# Geomorphometric atlas of ice-free Antarctic areas: Problem statement, concept, and key principles

## I.V. Florinsky[*]

Institute of Mathematical Problems of Biology, Keldysh Institute of Applied Mathematics, Russian Academy of Sciences, Pushchino, Moscow Region, 142290, Russia

**Abstract**

In this article, we rationalize the need to develop a geomorphometric atlas of ice-free Antarctic areas. Geomorphometric mapping of such landscapes is necessary to obtain new knowledge about quantitative topographic characteristics of these unique objects and the further use of morphometric information in the geosciences. First, we present a concept, structure, and content of the atlas including a list of 194 areas to be modeled and mapped. For each area, the atlas will include the following materials: (a) a hypsometric map; (b) a map series of the eleven morphometric variables: slope, aspect, horizontal curvature, vertical curvature, minimal curvature, maximal curvature, catchment area, topographic wetness index, stream power index, total insolation, and wind exposition index; and (c) reference texts and tables. Second, we describe data and methods of geomorphometric modeling and mapping to be utilized for creating the atlas. Fragments of the Reference Elevation Model of Antarctica will be used as input data for geomorphometric calculations and modeling. Next, we discuss specificity of the ice-free areas posing challenges in the atlas development. Finally, we consider possible applications of the derived maps. The principal feature and novelty of the atlas will be the system view on maps of key fundamental morphometric attributes associated with the theory of the topographic surface and the concept of general geomorphometry. The atlas will concentrate multi-scale, multi-aspect quantitative information on the ice-free Antarctic topography, will present it in a systematized, organized, and easy-to-read form as well as will provide scientific and information support for research in Antarctica.

## 1. Introduction

Ice-free, periglacial areas of Antarctica are usually classified as three main types of terrains (Markov et al. 1970; Simonov 1971; Korotkevich 1972; Alexandrov 1985; Pickard 1986; Beyer and Bölter 2002; Sokratova 2010):

1. Antarctic oases, i.e. coastal, shelf, and mountainous ice-free areas of Antarctica. Coastal oases are located between the continental ice sheet and the ocean; shelf oases are situated between the continental ice sheet and ice shelves; mountain oases can be observed in intramountain valleys and depressions.

2. Ice-free islands (or areas thereof) situated outside the ice shelves.

3. Ice-free mountain chains or their portions as well as nunataks.

Oases, ice-free mountain chains, and nunataks have the local climate of cold desert. The climate of the mountain ridges and nunataks is completely controlled by the surrounding ice sheet, while the climatic features of the oases—extremely contrasting, isolated intrazonal landscapes (Solopov 1973)—are determined by the mutual influence of adjacent glaciers and rocky surfaces of the oases (Solopov 1971). The climate of ice-free islands is affected by the interaction of the nearby continental ice sheet, ice shelves, and the ocean (Simonov 1975).

---

[*] Correspondence to: iflor@mail.ru

A special feature of oases and ice-free islands is the presence of a drainage networks including lakes and temporary summer streams (Markov et al. 1970; Simonov 1971, 1975; Korotkevich 1972; Alexandrov 1985).

In all types of ice-free areas of Antarctica, one can find primitive soils (Korotkevich 1972; Beyer and Bölter 2002; Bockheim 2015) and lower plants represented by algae, mosses, and lichens (Korotkevich 1972; Øvstedal and Lewis Smith 2001; Beyer and Bölter 2002; Ochyra et al. 2008). In addition to the soil fauna, some species of which can be observed even in very remote nunataks, the animal world of the shelf and mountain oases includes summer seabirds; pinnipeds and seabirds are widespread in the coastal oases and on the ice-free islands (Syroechkovsky 1966; Korotkevich 1972; Simonov 1975; Bulavintsev 1995; Beyer and Bölter 2002).

Scientists are still not completely clear on the mechanisms of the emergence and long-term stable self-sustaining existence of ice-free areas in the zone of the largest glaciation on the Earth (Model 1973; Sokratova 2010). Early hypotheses about increased heat flow have not been confirmed for such terrains, except for those located in regions of active volcanism. For the vast majority of the coastal and shelf oases, it is possible to accept the 'orographic' hypothesis which was first briefly formulated by Apfel (1948) and then developed by Markov et al. (1970): In places of dissected bedrock topography, an accelerated gravity-driven outflow of ice into the ocean occurs along subglacial valleys. Over elevated portions of glacier beds, where glacier movement velocities are decelerated, ablation does not have time to be compensated by the ice influx. As a result, bedrocks are exposed. Their thermal regime differs greatly from the regime of the snow and ice surface that leads to their increased heating and melting of the snow and ice around. Low air humidity promotes evaporation of moisture and melting of ice and snow, especially on the northern slopes. In the resulting oases, the mass balance is, on average, close to zero. This is because snow is blown by the wind into the sea, while melting ice has a lower albedo than snow (representative median values of albedo for snow-covered ice and melting glacier ice are 0.85 and 0.35, respectively (Grenfell 2011)) that leads to increased melting in summer. The ocean effect of bringing heat is also important for maintaining the existence of coastal and insular ice-free terrains. Climate-change induced deglaciation leads to a gradual increase in their size.

According to various estimates, the total area of ice-free rock outcrops ranges from 21,745 km² to 44,900 km², that is, from 0.18% to 0.37% of the total land area of Antarctica (Burton-Johnson et al. 2016a) (Fig. 1). Despite their relatively small size, these terrains are of great scientific and practical importance. Due to their relative accessibility, oases are convenient for the construction of year-round polar stations and seasonal field bases; most scientific work in austral summer is carried out in oases and islands (Sokratova 2010).

Topography is one of the most important components of the environment. Being the result of the interaction of endo- and exogenous processes at various scales and reflecting the geological structure, topography determines prerequisites for the gravity-driven migration and accumulation of moisture and other substances along the surface and in the near-surface layer, controls thermal, hydrological, and wind regimes, distribution of soil and vegetation cover, etc. The rigor and reproducibility of topography-oriented studies is currently ensured by geomorphometry, a science with a developed physical and mathematical theory and a powerful set of computational methods, the subject of which is modeling and analysis of topography as well as relationships between topography and other components of geosystems. Geomorphometric methods are widely used to solve various multiscale problems in the geosciences (Evans 1980; Moore et al. 1991; Wilson and Gallant 2000; Shary et al. 2002; Hengl and Reuter 2009; Minár et al. 2016; Wilson 2018; Florinsky 2017, 2025a).

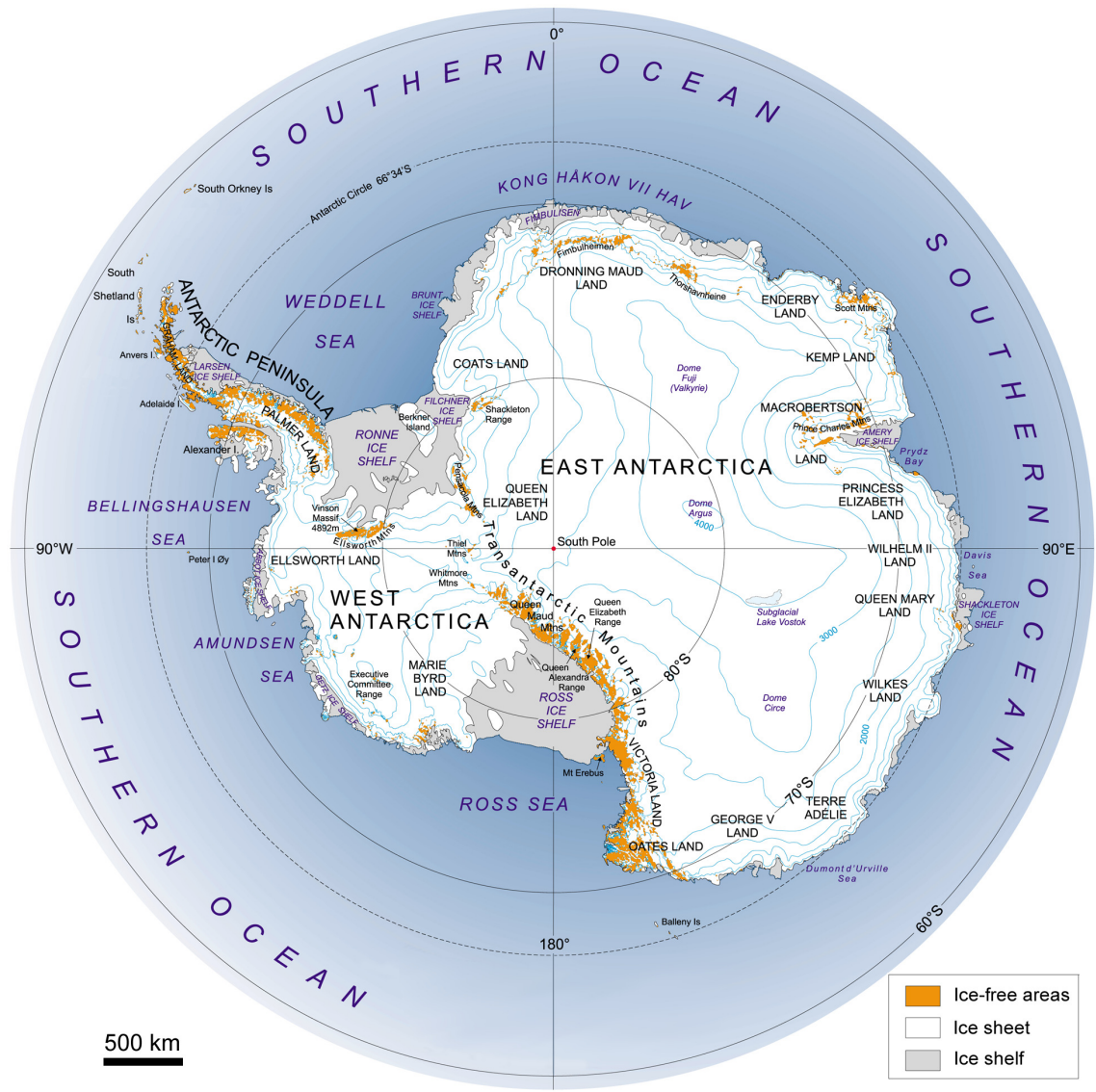

**Fig. 1** Ice-free areas of Antarctica. Modified from (BAS 2016)

Geomorphometric modeling and mapping of periglacial areas of the southern polar region is promising both for obtaining new knowledge about the quantitative characteristics of the topography of these unique objects, and for the further use of morphometric data for solving fundamental and applied problems of geomorphology, soil science, ecology, geology, geophysics, glaciology, climatology, and other geosciences.

Although the morphometry of Antarctica has been studied since the early 1960s, those works did not go beyond small-scale manual cartometric measurements of the continent perimeter and area, average heights of the ice sheet and subglacial surfaces, as well as the volume and thickness of the ice sheet (Suietova 1968). Later, Lastochkin (2006, 2007) has compiled a series of the small-scale morphometric maps of subglacial and submarine topography of Antarctica.

For some periglacial areas of the South Shetland Islands and the northern Antarctic Peninsula, several large-scale geomorphological and soil case studies had been conducted; those studies included the use of selected modern geomorphometric methods (Dąbski et al.

2017; Perondi et al. 2022; Siqueira et al. 2022, 2023). For ten ice-free Antarctic areas—the Larsemann Hills, Thala Hills (Molodezhny and Vecherny Oases), Schirmacher Hills, Fildes Peninsula, Bunger Hills, Cape Burks, Haswell Island, Leningradsky Nunatak, and Gaussberg Volcano—we performed geomorphometric modeling and mapping (Florinsky 2023a, 2023b, 2025b, 2025c; Florinsky and Zharnova 2025). However, geomorphometric mapping of such terrains needs to be posed on a larger scope and more comprehensively. In particular, it would be expedient to create a geomorphometric atlas of all ice-free areas of Antarctica. The creation of such an atlas will allow generalizing and systematizing regional morphometric information in a form traditional for the geosciences.

Over the years of mapping the southern polar region (Aleiner 1950; Koblenz 1964; Clancy et al. 2014), several comprehensive physical geographical, scientific reference atlases of Antarctica have been created. Today some of them are of only historical interest (e.g., Bakaev and Tolstikov 1966; SHO 1993). There is a relatively current fundamental atlas of Antarctica (Korotkevich et al. 2005), but the most of its maps are presented in small scales and geomorphometric information is missing.

There are three thematic physical geographical, scientific reference atlases of Antarctica. A series of small-scale morphological maps of subglacial and submarine topography as well as traditional geomorphological maps for some oases can be found in a regional geomorphological atlas (Lastochkin 2011), which implemented a controversial morphodynamic concept of its editor (Lastochkin 2006, 2007, 2018). Two other atlases present: (a) small-scale maps of the Antarctic topography obtained on the basis of geostatistical analysis of satellite radar altimetry data (Herzfeld 2004) and (b) small-scale maps of the Antarctic gravity field (Klokočník et al. 2017). There are no geomorphometric components in these three atlases.

In addition to printed atlases, there are several pan-Antarctic geoinformation products, which can be considered as website or desktop alternatives to traditional atlases. For example, the US Geological Survey has created a general geographic geoportal called Antarctic Research Atlas (USGS 1999–2024). This geoportal allows online visualization of several information layers, such as geographic names, hydrography, elevations, satellite imagery, and outlines of rock outcrops (that is, ice-free areas). GeoMAP geoportal depicts online a detailed geological map of Antarctic exposed bedrock and surficial geology (Cox et al. 2023). One can download and use two pan-Antarctic desktop datasets on (a) the mean annual ground temperatures of the permafrost (Obu et al. 2020) and (b) the hierarchical classification and mapping of ice-free land ecosystems (Tóth et al. 2025). However, there are no geomorphometric components in these four geoinformation products (although Tóth et al. (2025) used some morphometric measures (i.e., slope and aspect) for the ecosystem classification).

The only pan-Antarctic geoinformation product providing some limited morphometric information is the REMA Explorer (PGC 2022–2024), which handles online Reference Elevation Model of Antarctica (REMA) (Howat et al. 2019) (Section 2.2.1). The REMA Explorer tools include visualization of two morphometric layers derived from REMA—slope gradient and aspect—for the entire territory of Antarctica (Summer 2020). However, first, the REMA Explorer does not allow searching or identifying ice-free areas and, second, the use of two indicated morphometric variables are not enough for conducting comprehensive scientific research in the second quarter of the 21st century.

The purpose of this article is to rationalize the need to develop a geomorphometric atlas of ice-free Antarctic areas, to present a concept, structure, and content of the atlas, to list terrains to be mapped, as well as to describe data and methods of geomorphometric modeling and mapping to be utilized for creating the atlas.

# 2. Materials and methods
## 2.1. Atlas concept, structure, and content
### 2.1.1. Concept

It is well known that an atlas is not just a set of various maps. On the contrary, an atlas is a system of maps, which are interrelated to one another and complement each other (Salichtchev 1990, p. 185). The goal of our project is to develop a physical geographical, scientific reference geomorphometric atlas of ice-free Antarctic areas. The principal attribute of this atlas is that we develop a system of maps of key fundamental morphometric variables (Sections 2.1.4 and 2.2.3), which are associated with the theory of the topographic surface (Shary et al. 2002; Florinsky 2025a) and the concept of general geomorphometry. The latest is defined as "the measurement and analysis of those characteristics of landform which are applicable to any continuous rough surface. ... General geomorphometry … provides a basis for the quantitative comparison … of qualitatively different landscapes …." (Evans 1972, p. 18). Each morphometric variable describes a certain property of the topography and has a unique physical mathematical and physical geographical interpretation; various morphometric variables complement each other (Florinsky 2025a).

### 2.1.2. Structure

The atlas will include the following chapters:
- An introductory chapter.
- 20 regional chapters generally organized by Antarctic lands (main geographical units of Antarctica), from the prime meridian clockwise: (1) Queen Maud Land, (2) Enderby Land, (3) Kemp Land, (4) MacRobertson Land, (5) Princess Elizabeth Land, (6) Wilhelm II Land, (7) Queen Mary Land, (8) Wilkes Land, (9) George V Land, (10) Oates Land, (11) Victoria Land, (12) Central Transantarctic Mountains, (13) Marie Byrd Land, (14) Ellsworth Land, (15) Alexander I Island, (16) Palmer Land, (17) Graham Land, (18) South Shetland Islands, (19) Queen Elizabeth Land, and (20) Coats Land.
- Reference tables.
- References.
- Index of geographic names.

### 2.1.3. Ice-free areas

All ice-free areas will be grouped within regional chapters. Table 1 includes 194 completely or partially ice-free Antarctic areas, which will be presented in the atlas. We compiled this list by analyzing available topographic maps (Bakaev and Tolstikov 1966; USGS 1959–1973; Korotkevich et al. 2005; AAD 2023), satellite image mosaics from the REMA Explorer (PGC 2022–2024), Landsat-8 derived rock outcrop dataset (Burton-Johnson et al. 2016b), as well as information presented in the Antarctic Research Atlas (USGS 1999–2024).

Compiling the list, we checked completely and partially ice-free areas of Antarctica according to the criterion of their size. This is because in geomorphometric modeling and mapping, a limiting factor is a grid spacing of the input data, digital elevation models (DEMs) (Florinsky 2025a, chap. 3). The most detailed set of the input data, REMA, has a grid spacing of 2 m (see Section 2.2.1). Therefore, it makes no sense to include in the list small nunataks and coastal cliffs with planimetric sizes less than 200–300 m, because their morphometric maps will be too rasterized, unreadable, and useless.

The names of ice-free areas (Table 1) were determined and checked using topographic maps of Antarctica at scales of 1: 1,000,000, 1: 250,000, 1: 50,000, and 1: 25,000 (Bakaev and Tolstikov 1966; USGS 1959–1973; Korotkevich et al. 2005; AAD 2023), the Antarctic Research Atlas (USGS 1999–2024) as well as several gazetteers (Donidze 1987; Alberts 1995;

**Table 1 Ice-free Antarctic areas**

| Satellite imagery map and area numbers | Area name | Geographical coordinates |
|---|---|---|
| **I** | *Queen Maud Land* | |
| 1 | Gburek Peaks | 72.19071° S, 0.33728° W |
| 2 | Sverdrup Mountains | 72.49954° S, 0.86334° E |
| 3 | Gjelsvik Mountains | 72.16798° S, 2.79840° E |
| 4 | Mühlig-Hofmann Mountains | 72.02310° S, 5.58629° E |
| 5 | Filchner Mountains | 72.00842° S, 7.51696° E |
| 6 | Kurze Mountains | 71.87007° S, 8.89116° E |
| 7 | Conrad Mountains | 71.84385° S, 9.69159° E |
| 8 | Humboldt Mountains | 71.73975° S, 11.49972° E |
| 9 | Wohlthat Mountains | 71.64117° S, 12.25986° E |
| 10 | Untersee Oasis | 71.34897° S, 13.45676° E |
| 11 | Weyprecht Mountains | 72.00571° S, 13.39693° E |
| 12 | Payer Mountains | 71.98383° S, 14.58366° E |
| 13 | Lomonosov Mountains | 71.49440° S, 15.47366° E |
| 14 | Schirmacher Hills | 70.75863° S, 11.63975° E |
| **II** | *Queen Maud Land* | |
| 15 | Sør Rondane Mountains | 72.07342° S, 25.12917° E |
| 16 | Belgica Mountains | 72.60218° S, 31.28041° E |
| 17 | Queen Fabiola Mountains | 71.54932° S, 35.68163° E |
| **III** | *Queen Maud Land – Enderby Land* | |
| 18 | Rundvågs Hills | 69.91214° S, 39.02842° E |
| 19 | Skallen Hills | 69.66762° S, 39.43763° E |
| 20 | Skarvsnes Foreland | 69.47392° S, 39.68363° E |
| 21 | Langhovde Hills | 69.22389° S, 39.69693° E |
| 22 | Flatvaer Islands | 69.02852° S, 39.54379° E |
| 23 | Cape Hinode | 68.15075° S, 42.68939° E |
| 24 | Polkanov Hills | 67.97402° S, 44.06999° E |
| 25 | Tereshkova Oasis | 67.95424° S, 44.55420° E |
| 26 | Konovalov Mountains | 67.75899° S, 45.76434° E |
| 27 | Thala Hills – Molodezhny Oasis | 67.66987° S, 45.87278° E |
| 28 | Thala Hills – Vecherny Oasis | 67.65804° S, 46.11360° E |
| **IV** | *Enderby Land* | |
| 29 | Berge der Deutsch-Sowjetischen Freundschaft | 67.98680° S, 47.36288° E |
| 30 | Nye Mountains | 68.14730° S, 49.04672° E |
| 31 | Raggatt Mountains | 67.72608° S, 49.12536° E |
| 32 | Fyfe Hills | 67.36599° S, 49.23295° E |
| 33 | Scott Mountains | 67.49827° S, 50.46000° E |
| 34 | Tula Mountains | 66.95319° S, 51.32388° E |
| 35 | Napier Mountains | 66.51256° S, 53.64992° E |
| **V** | *Kemp Land – MacRobertson Land* | |
| 36 | Leckie Range | 67.91636° S, 56.46961° E |
| 37 | Kvars Promontory | 67.03474° S, 57.03544° E |
| 38 | Øygarden Group | 66.96244° S, 57.56056° E |
| 39 | Stillwell Hills | 67.39222° S, 59.44065° E |
| 40 | Stump Mountain | 67.46772° S, 60.97262° E |
| 41 | Ufs Island | 67.48211° S, 61.13982° E |
| 42 | David Range | 67.85738° S, 62.52450° E |
| 43 | Masson Range | 67.84066° S, 62.84619° E |
| **VI** | *MacRobertson Land – Princess Elizabeth Land* | |
| 44 | Prince Charles Mountains | 70.78079° S, 66.35888° E |
| 45 | Jetty Oasis | 70.81579° S, 68.12083° E |
| 46 | Blake Nunataks | 74.23840° S, 67.11461° E |
| 47 | Cumpston Massif | 73.61873° S, 66.75461° E |
| 48 | Mawson Escarpment | 73.10501° S, 68.21482° E |
| 49 | Clemence Massif | 72.19473° S, 68.63958° E |
| 50 | Reinbolt Hills | 70.48101° S, 72.53746° E |
| 51 | Mistichelli Hills | 70.03499° S, 72.83590° E |
| 52 | Landing Bluff | 69.74317° S, 73.71167° E |
| 53 | Grove Mountains | 72.89365° S, 75.05047° E |
| **VII** | *Princess Elizabeth Land* | |
| 54 | Bølingen Islands | 69.45033° S, 75.71897° E |
| 55 | Larsemann Hills | 69.39450° S, 76.31772° E |
| 56 | Rauer Group | 68.83439° S, 77.81831° E |
| 57 | Vestfold Hills | 68.57153° S, 78.17928° E |
| **VIII** | *Wilhelm II Land – Queen Mary Land* | |
| 58 | Gaussberg | 66.80408° S, 89.19631° E |
| 59 | Haswell Islands | 66.52617° S, 92.99577° E |
| **IX** | *Queen Mary Land – Wilkes Land* | |
| 60 | Obruchev Hills | 66.56093° S, 99.84070° E |
| 61 | Highjump Archipelago | 66.06099° S, 100.66434° E |
| 62 | Bunger Hills | 66.26764° S, 100.86504° E |
| 63 | Browning Peninsula | 66.46909° S, 110.54866° E |
| 64 | Bailey Peninsula | 66.29294° S, 110.55327° E |
| **X** | *George V Land* | |
| 65 | Cape Denison | 67.00904° S, 142.66538° E |
| 66 | Horn Bluff | 68.38860° S, 149.78208° E |
| 67 | SCAR Bluffs | 68.81302° S, 153.53338° E |
| **XI** | *Oates Land* | |
| 68 | Lazarev Mountains | 69.44490° S, 157.18452° E |
| 69 | Wilson Hills | 69.55913° S, 158.30665° E |
| 70 | Holladay Nunataks | 69.53032° S, 159.32131° E |
| **XII** | *Oates Land – Victoria Land* | |
| 71 | USARP Mountains | 71.28672° S, 160.26706° E |
| 72 | Bowers Mountains | 71.19928° S, 163.18917° E |
| 73 | Concord Mountains | 71.25353° S, 166.91001° E |
| 74 | Freyberg Mountains | 72.26214° S, 163.55320° E |
| 75 | Admiralty Mountains | 71.70077° S, 169.35384° E |
| 76 | Victory Mountains | 72.67712° S, 168.43169° E |
| 77 | Mesa Range | 73.22919° S, 162.81079° E |
| 78 | Deep Freeze Range | 74.23594° S, 163.58338° E |
| 79 | Eisenhower Range | 74.36801° S, 162.56545° E |
| **XIII** | *Victoria Land* | |
| 80 | Ricker Hills | 75.68314° S, 159.25865° E |
| 81 | Prince Albert Mountains | 76.76219° S, 160.76997° E |
| 82 | McMurdo Dry Valleys | 77.45272° S, 161.72625° E |
| 83 | Royal Society Range | 78.05319° S, 162.86601° E |
| 84 | Denton Hills | 78.12376° S, 163.97833° E |
| 85 | Ross Island | 77.54762° S, 167.16211° E |
| 86 | Mount Discovery | 78.34179° S, 165.23764° E |
| 87 | Warren Range | 78.39340° S, 158.32048° E |
| 88 | Boomerang Range | 78.56511° S, 158.78751° E |
| **XIV** | *Victoria Land* | |
| 89 | Cook Mountains | 79.47427° S, 158.48418° E |
| 90 | Darwin Mountains | 79.47427° S, 158.48418° E |
| 91 | Britannia Range | 80.17766° S, 157.78339° E |
| 92 | Churchill Mountains | 81.56919° S, 158.82292° E |
| 93 | Surveyors Range | 81.73777° S, 159.88110° E |
| 94 | Nash Range | 80.89228° S, 158.19967° E |
| 95 | Holyoake Range | 82.26022° S, 160.25738° E |
| 96 | Cobham Range | 82.35294° S, 159.21239° E |
| **XV** | *Central Transantarctic Mountains* | |
| 97 | Geologists Range | 82.47914° S, 155.85908° E |
| 98 | Miller Range | 83.09903° S, 156.94750° E |
| 99 | Queen Elizabeth Range | 83.32849° S, 161.55652° E |
| 100 | Holland Range | 83.18688° S, 166.02209° E |
| 101 | Queen Alexandra Range | 84.07468° S, 167.14550° E |
| 102 | Dominion Range | 85.33201° S, 166.94308° E |
| 103 | Queen Maud Mountains | 84.62216° S, 172.34636° E |
| 104 | Hughes Range | 84.38594° S, 174.54274° E |
| 105 | Prince Olav Mountains | 84.95509° S, 174.67127° E |
| 106 | Rawson Mountains | 86.47909° S, 156.06823° W |
| 107 | Horlick Mountains | 85.89615° S, 128.14237° W |
| **XVI** | *Marie Byrd Land* | |
| 108 | Fosdick Mountains | 76.50716° S, 145.20621° W |
| 109 | Ford Ranges | 76.96917° S, 145.13782° W |
| 110 | Erickson Bluffs | 75.02302° S, 136.67668° W |
| 111 | Cape Burks | 74.75846° S, 136.81514° W |
| 112 | Demas Range | 75.04530° S, 133.76415° W |
| 113 | Flood Range | 76.03680° S, 135.16703° W |
| 114 | Ames Range | 75.76550° S, 132.30298° W |
| 115 | McCuddin Mountains | 75.73630° S, 129.16838° W |
| 116 | Executive Committee Range | 77.07196° S, 125.91917° W |
| **XVII** | *Ellsworth Land* | |
| 117 | Hudson Mountains | 74.63586° S, 99.42615° W |
| **XVIII** | *Ellsworth Land* | |
| 118 | Thiel Mountains | 85.19731° S, 90.23627° W |
| 119 | Whitmore Mountains | 82.48111° S, 103.77870° W |
| 120 | Nash Hills | 81.92911° S, 89.61358° W |
| 121 | Pirrit Hills | 81.13601° S, 85.51966° W |
| 122 | Ellsworth Mountains | 78.91080° S, 84.56665° W |
| **XIX** | *Alexander I Island – Palmer Land* | |
| 123 | Debussy Heights | 69.90916° S, 71.31294° W |
| 124 | Hornpipe Heights | 69.87550° S, 70.58434° W |
| 125 | Lully Foothills | 70.79477° S, 69.63323° W |
| 126 | Staccato Peaks | 71.76149° S, 70.58631° W |
| 127 | Ablation Valley | 70.79918° S, 68.44435° W |
| 128 | Ganymede Heights | 70.88166° S, 68.41346° W |
| 129 | Elephant Ridge | 71.35204° S, 68.32555° W |
| 130 | Horrocks Block | 71.58709° S, 68.31923° W |
| 131 | Mars Oasis | 71.87370° S, 68.27098° W |
| 132 | Behrendt Mountains | 75.31101° S, 72.50996° W |
| 133 | Merrick Mountains | 75.08919° S, 72.05406° W |
| 134 | Sweeney Mountains | 75.15157° S, 69.41835° W |
| 135 | Hauberg Mountains | 75.15157° S, 69.41835° W |
| 136 | Scaife Mountains | 75.15870° S, 64.97666° W |
| 137 | Latady Mountains | 74.69710° S, 64.26198° W |
| 138 | Rare Range | 74.37914° S, 64.10655° W |
| 139 | Guettard Range | 74.40266° S, 63.18271° W |
| 140 | Hutton Mountains | 74.13159° S, 62.51864° W |
| 141 | Werner Mountains | 73.54753° S, 62.41937° W |
| 142 | Dana Mountains | 73.54753° S, 62.41392° W |
| 143 | Carey Range | 72.96578° S, 62.34251° W |
| 144 | Wegener Range | 72.72226° S, 62.26823° W |
| 145 | Peck Range | 72.22468° S, 62.42147° W |
| 146 | Hess Mountains | 71.98978° S, 62.51799° W |
| 147 | Schirmacher Massif | 71.59502° S, 62.28781° W |
| 147 | Cat Ridge | 71.08915° S, 61.94099° W |
| 148 | Parmelee Massif | 70.97531° S, 62.12792° W |
| 149 | Welch Mountains | 70.93036° S, 63.45534° W |
| 150 | Eland Mountains | 70.50707° S, 62.99098° W |
| 151 | Batterbee Mountains | 71.47131° S, 67.26160° W |
| 152 | Canis Heights | 70.43027° S, 66.38876° W |
| 153 | Campbell Ridges | 70.38082° S, 67.50575° W |
| 154 | Traverse Mountains | 69.91887° S, 67.94935° W |
| **XX** | *Graham Land* | |
| 155 | Square Bay Islands | 67.76182° S, 67.00835° W |
| 156 | Kennet Mounts | 67.20999° S, 65.18534° W |
| 157 | McClary Ridge | 66.82915° S, 64.18921° W |
| 158 | Bigla Ridge | 66.53701° S, 63.91961° W |
| 159 | Cape Casey | 66.36665° S, 63.71748° W |
| 160 | Lyttelton Ridge | 66.28496° S, 63.11276° W |
| 161 | Voden Heights | 65.83951° S, 62.86384° W |
| 162 | Aristotle Mountains | 65.54510° S, 62.48522° W |
| 163 | Austa Ridge | 65.24498° S, 62.19916° W |
| 164 | Metlichina Ridge | 65.11417° S, 61.93080° W |
| 165 | Rugate Ridge | 65.01824° S, 61.94326° W |
| 166 | Zagreus Ridge | 64.87087° S, 61.84613° W |
| 167 | Tillberg Ridge | 64.79962° S, 60.97322° W |
| 168 | Nordenskjöld Coast | 64.50060° S, 60.34802° W |
| 169 | Eagle Island | 63.65722° S, 57.49492° W |
| 170 | Vega Island | 63.84113° S, 57.42613° W |
| 171 | Seymour Island | 64.28021° S, 56.76862° W |
| 172 | James Ross Island | 64.11479° S, 57.95821° W |
| **XXI** | *South Shetland Islands* | |
| 173 | Deception Island | 62.94944° S, 60.64135° W |
| 174 | Byers Peninsula and Rugged Island | 62.64061° S, 61.07264° W |
| 175 | Cape Shirreff | 62.46768° S, 60.78828° W |
| 176 | Hurd Peninsula | 62.70615° S, 60.40990° W |
| 177 | Rozhen Peninsula | 62.75455° S, 60.31997° W |
| 178 | Fildes Peninsula and Ardley Island | 62.19073° S, 58.96010° W |
| 179 | Barton Peninsula | 62.22833° S, 58.74738° W |
| 180 | Potter Peninsula | 62.24743° S, 58.65523° W |
| 181 | Western coast of Admiralty Bay | 62.18541° S, 58.44579° W |
| 182 | Keller Peninsula | 62.07495° S, 58.40609° W |
| **XXII** | *Queen Elizabeth Land* | |
| 183 | Patuxent Range | 84.62679° S, 63.85130° W |
| 184 | Neptune Range | 83.87829° S, 56.44364° W |
| 185 | Pensacola Mountains | 83.62408° S, 55.31863° W |
| 186 | Forrestal Range | 83.19049° S, 50.63544° W |
| 187 | Dufek Massif | 82.54255° S, 51.52901° W |
| 188 | Argentina Range | 82.31155° S, 42.15123° W |
| **XXIII** | *Coats Land* | |
| 189 | Shackleton Range | 80.51287° S, 27.08618° W |
| **XXIV** | *Queen Maud Land* | |
| 190 | Kraul Mountains | 73.40679° S, 14.14385° W |

| 191 | Heimefront Range | 80.51287° S, 27.08618° W |
| 192 | Kirwan Escarpment | 73.89704° S, 5.13154° W |
| 193 | Borg Massif | 73.89704° S, 5.13154° W |
| 194 | Ahlmann Ridge | 72.11798° S, 2.90514° W |

SCAR 1992–2025). Geographic coordinates of the conditional centers of the ice-free areas (Table 1) were determined in the WGS-1984 coordinate system using the REMA Explorer (PGC 2022–2024).

For a clear understanding of the spatial distribution and configuration of the ice-free Antarctic areas, we compiled a series of 24 satellite imagery maps with the designation of all selected objects using the REMA Explorer (PGC 2022–2024). A layout of satellite imagery maps is shown in Fig. 2. For a series of satellite imagery maps, see Fig. 3.

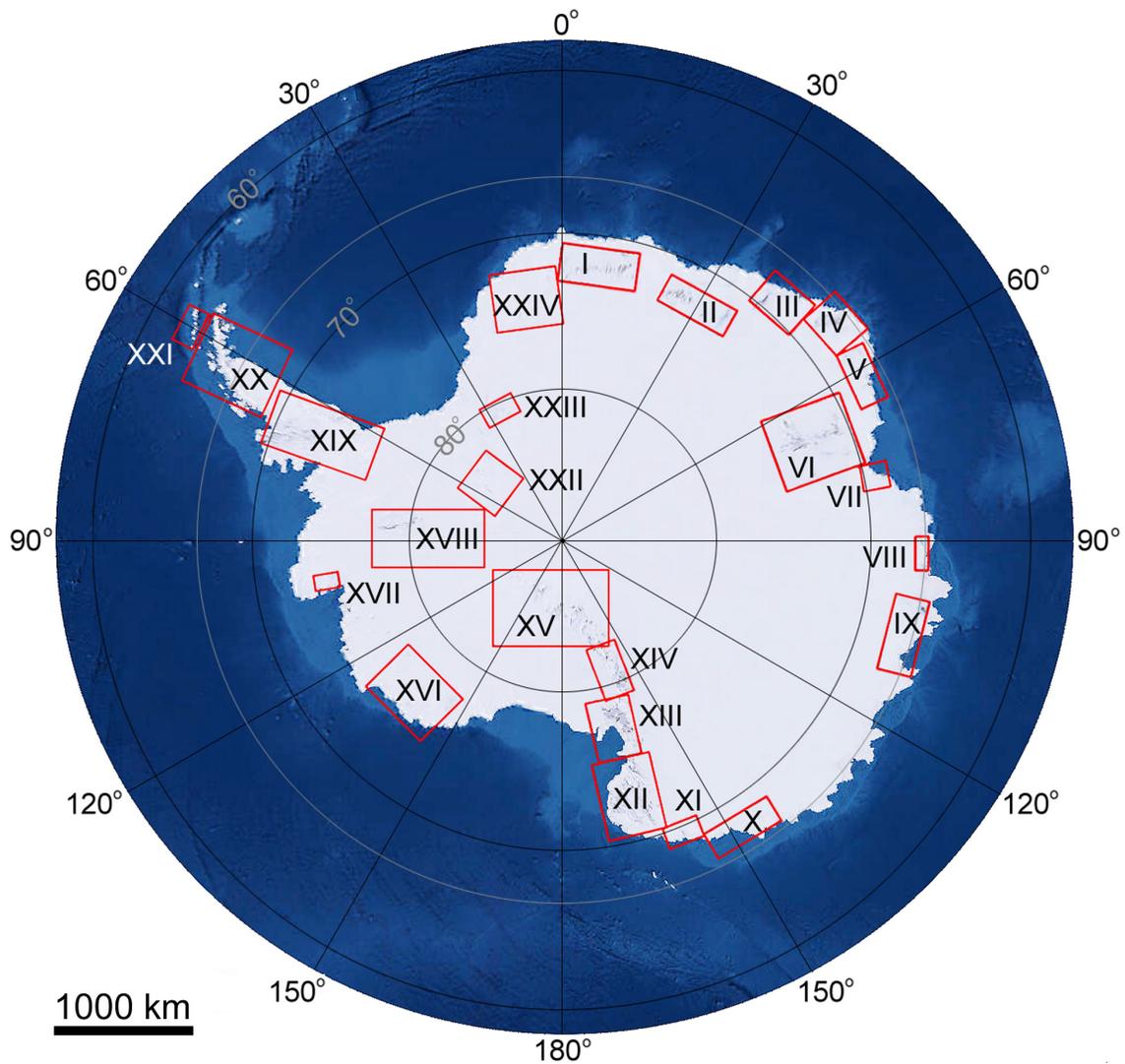

**Fig. 2** Layout of satellite imagery maps (red rectangles, see Fig. 3) with ice-free areas of Antarctica. Latin numerals are satellite imagery map numbers. Background is from REMA Explorer (PGC 2022–2024)

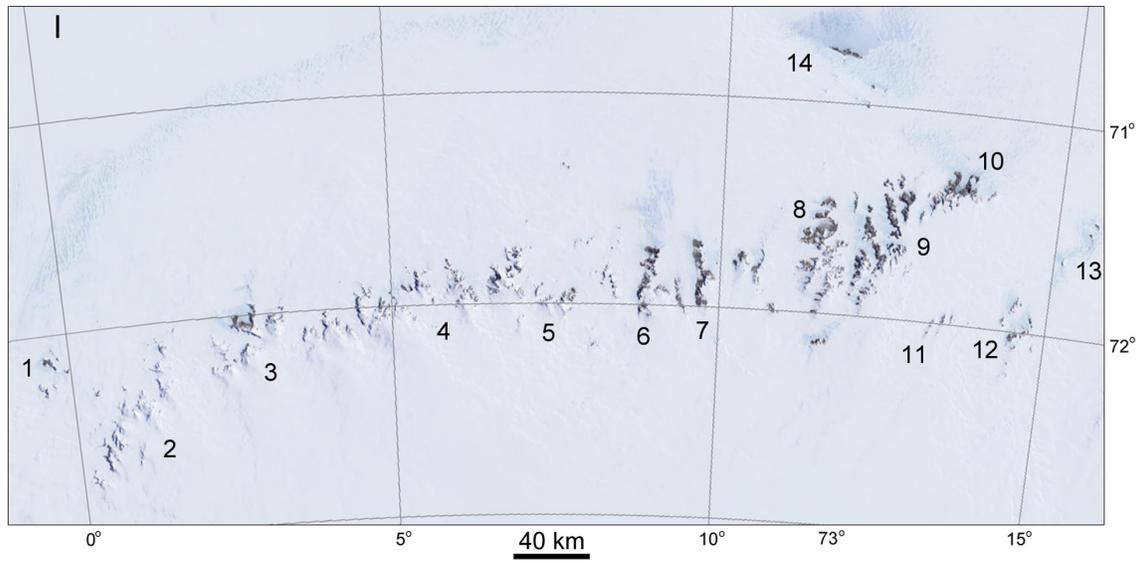

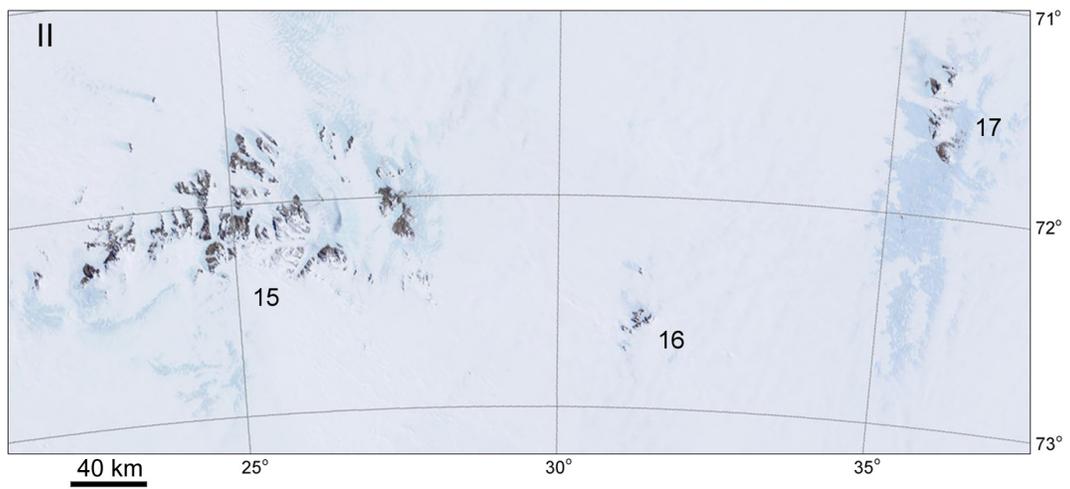

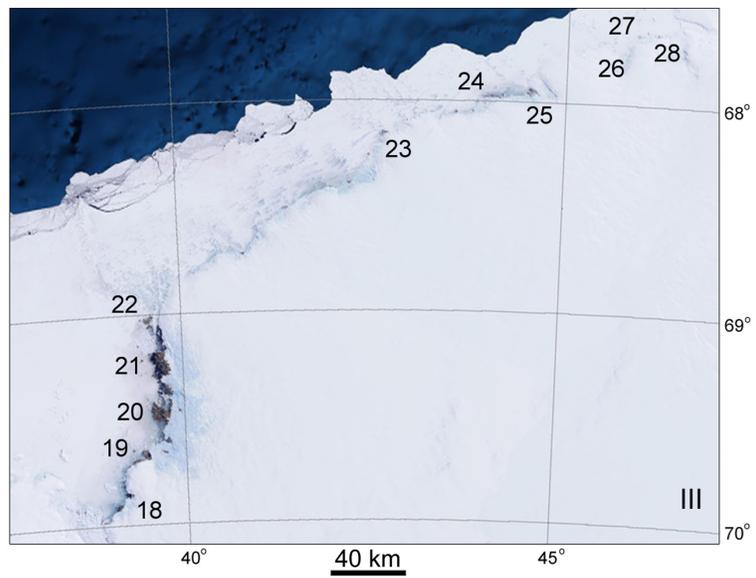

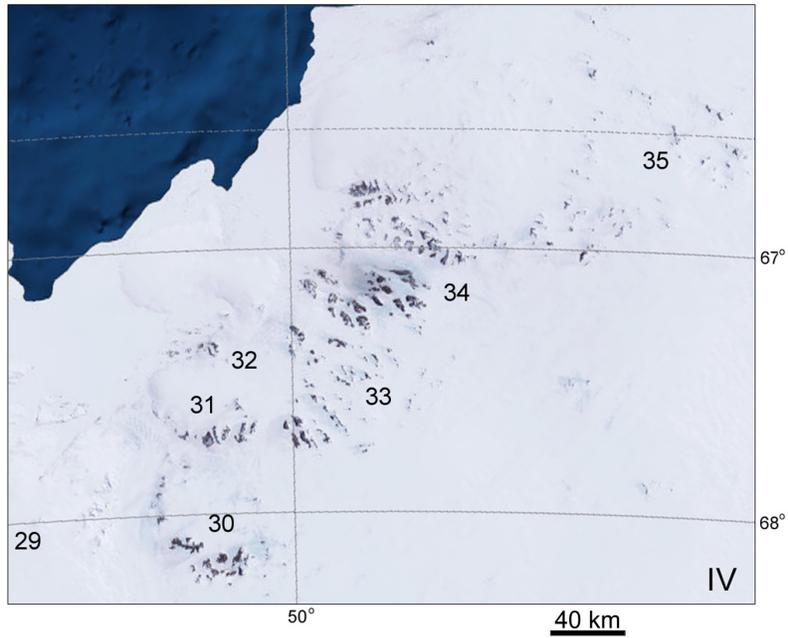

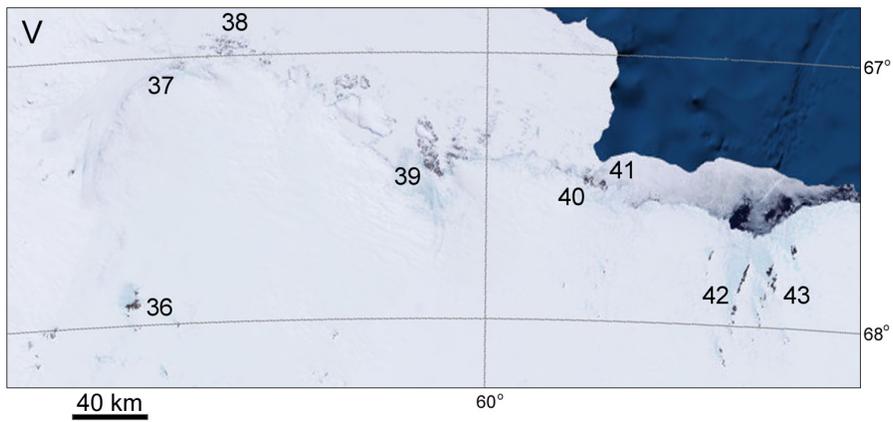

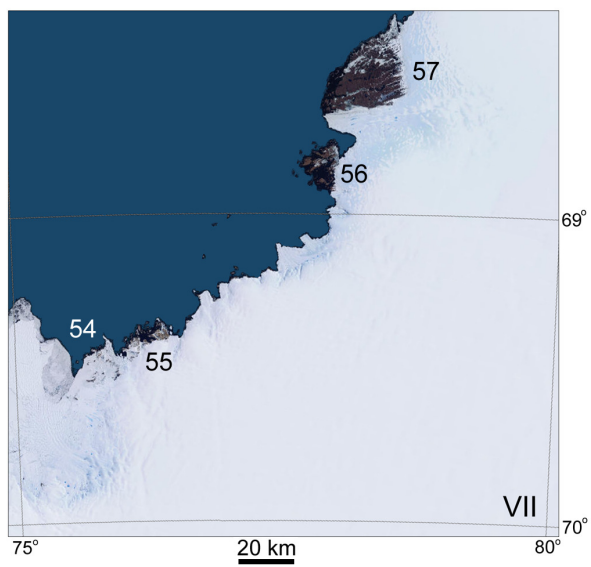

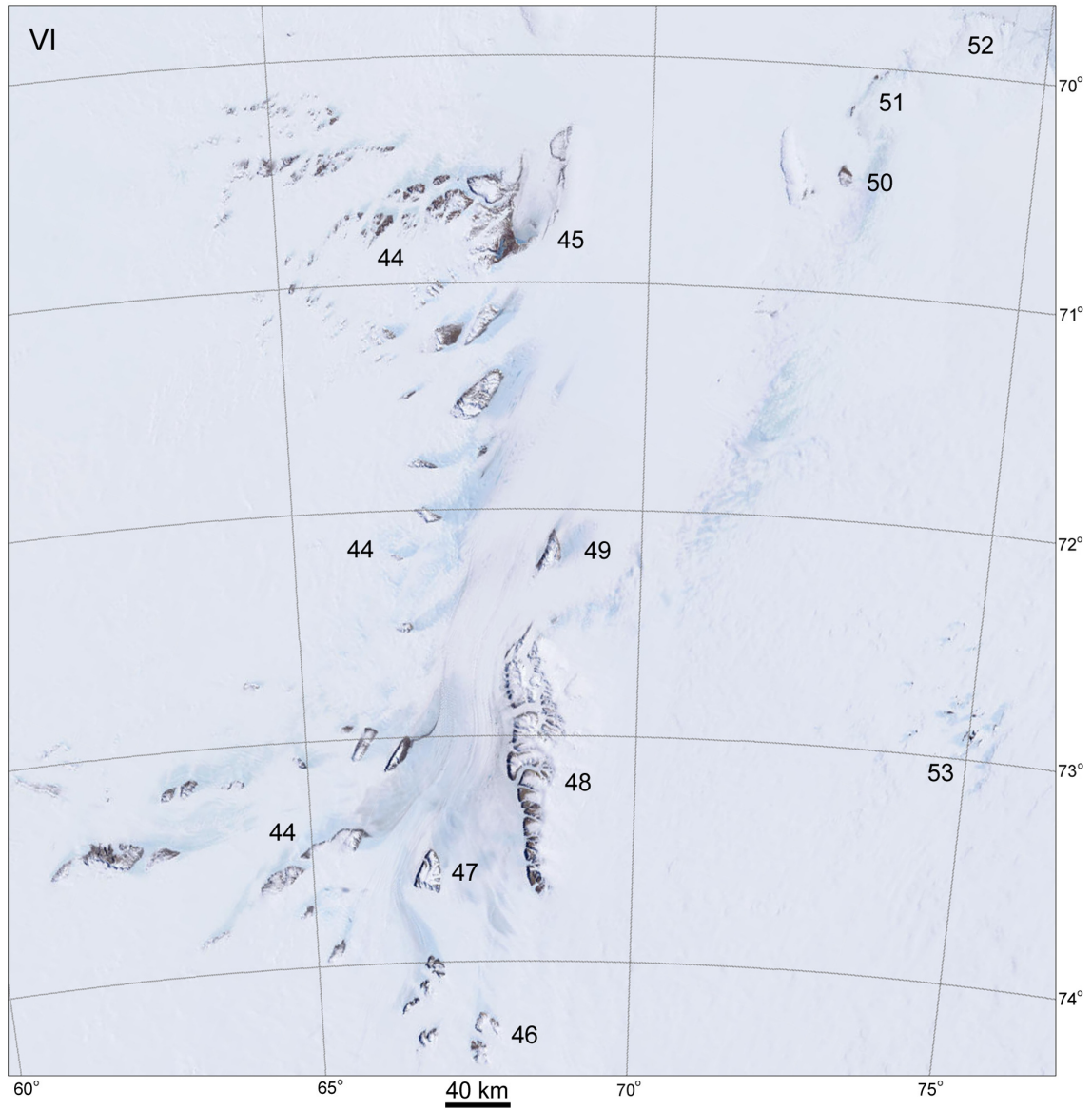

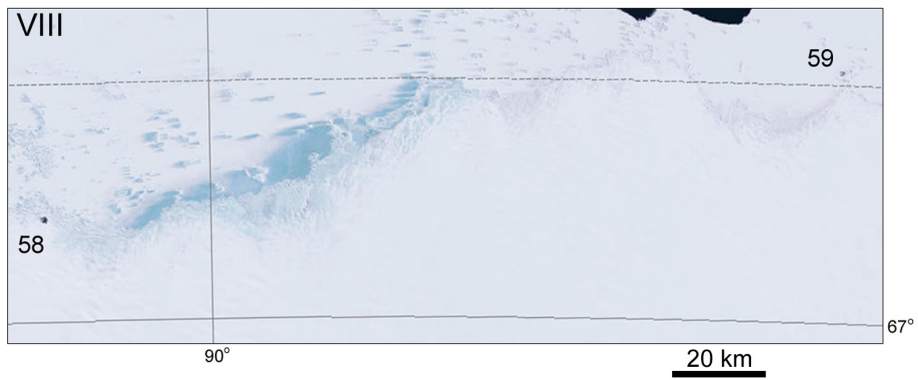

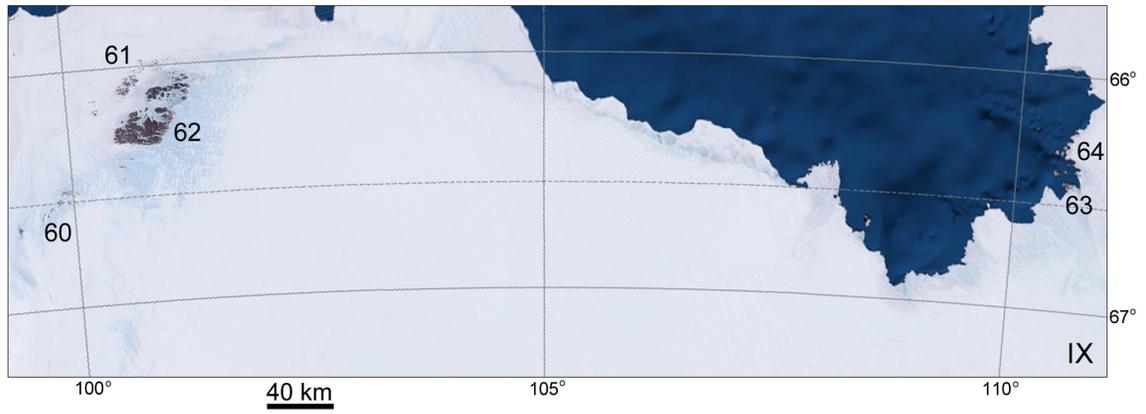

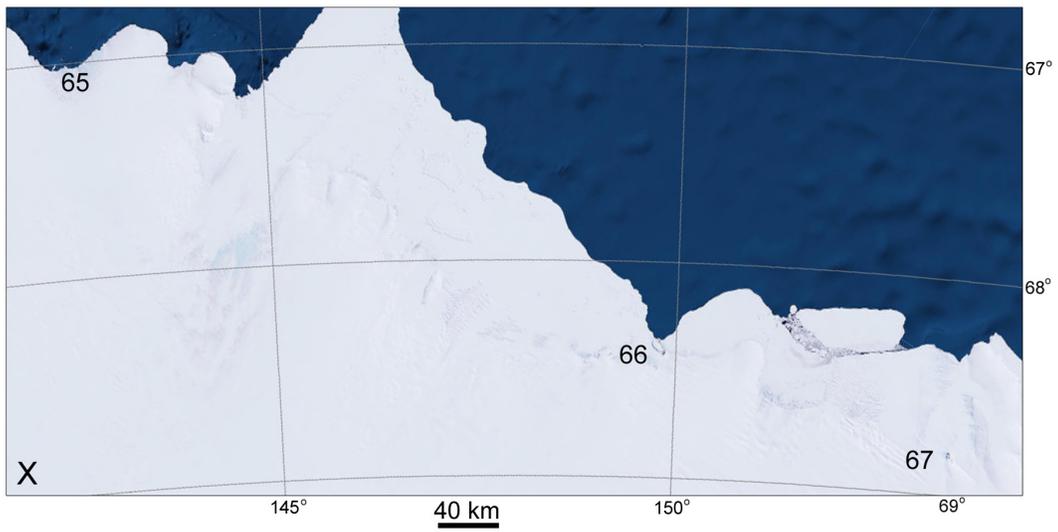

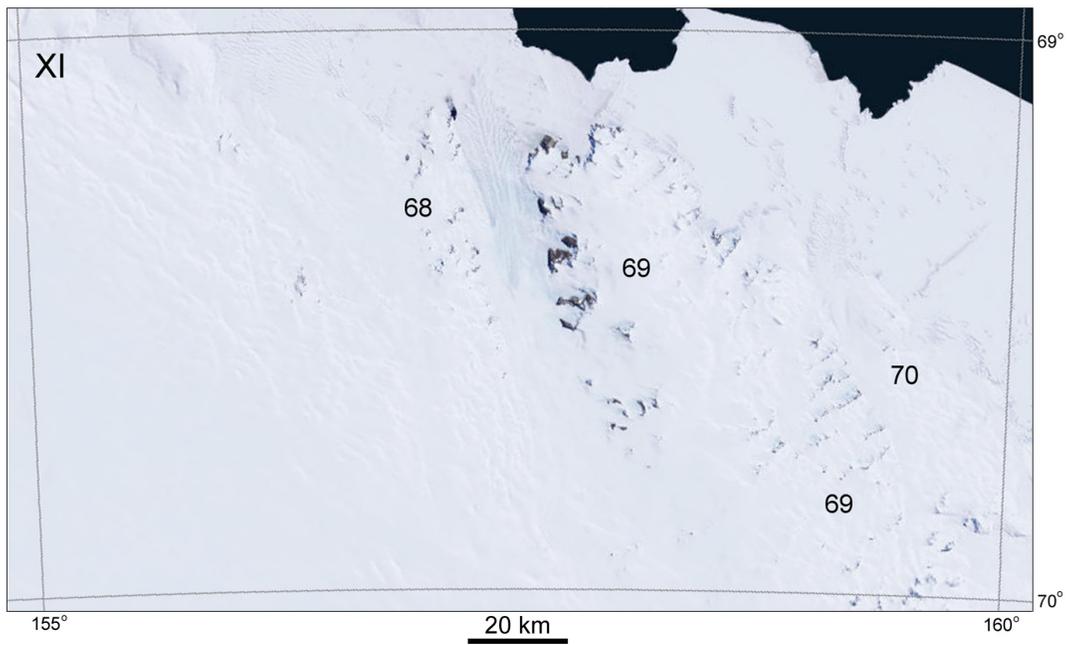

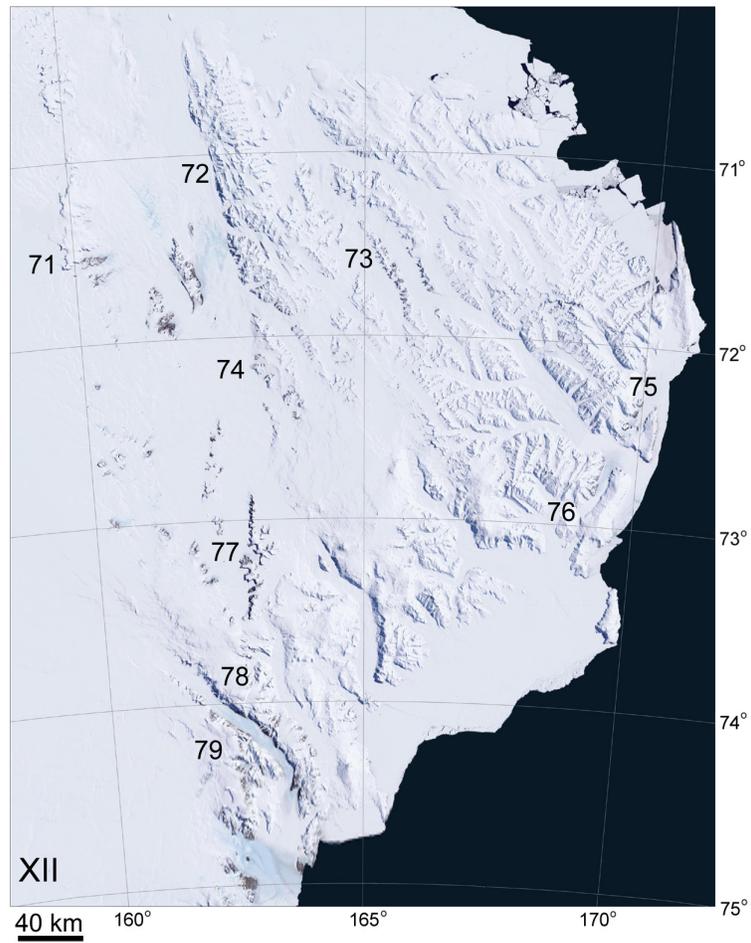

XII

40 km  160°    165°    170°

71°

72°

73°

74°

75°

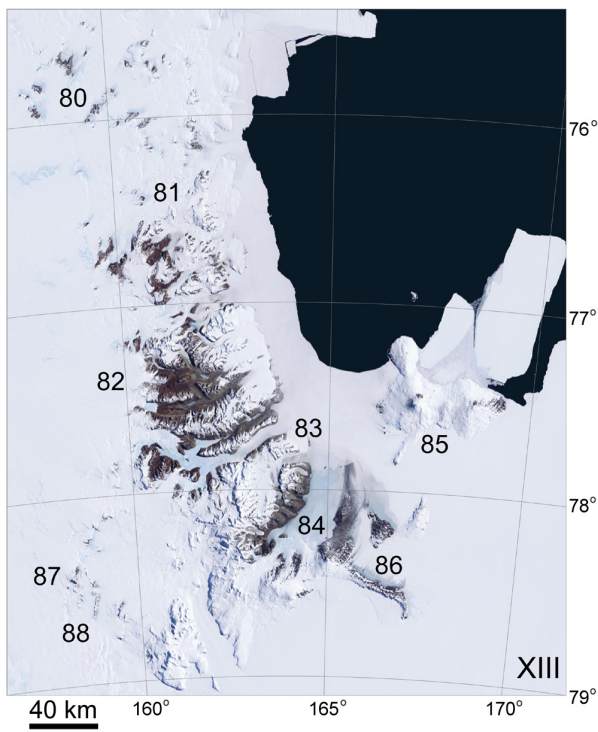

XIII

40 km  160°    165°    170°

76°

77°

78°

79°

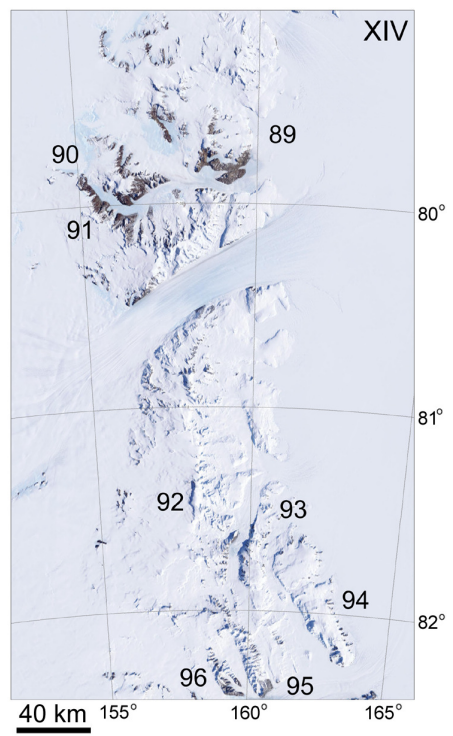

XIV

40 km  155°    160°    165°

80°

81°

82°

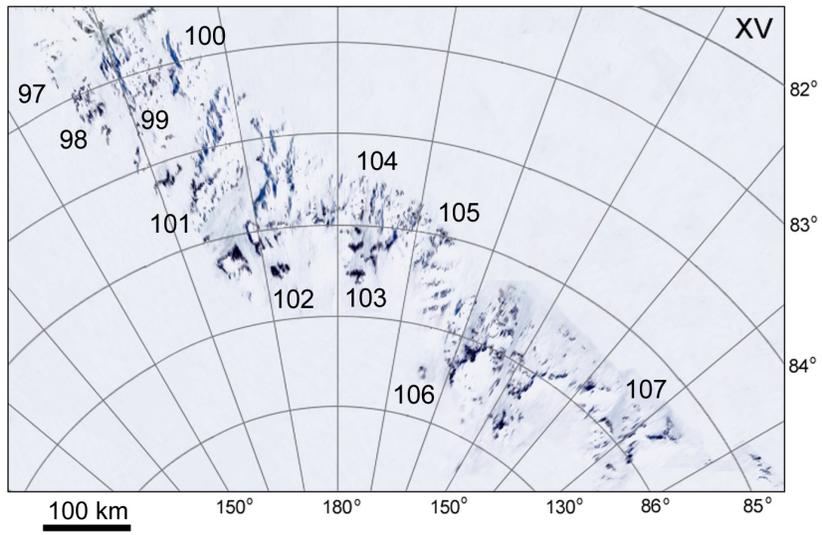

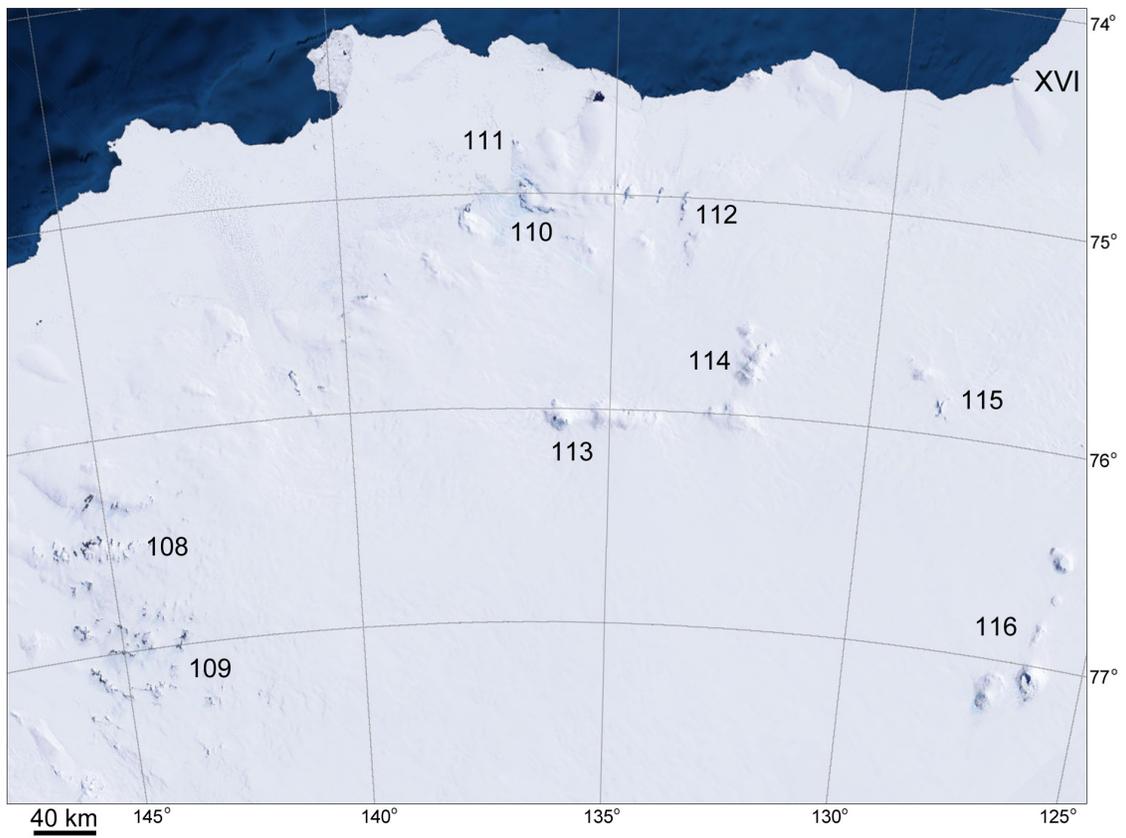

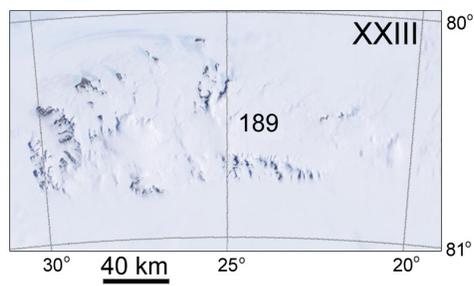

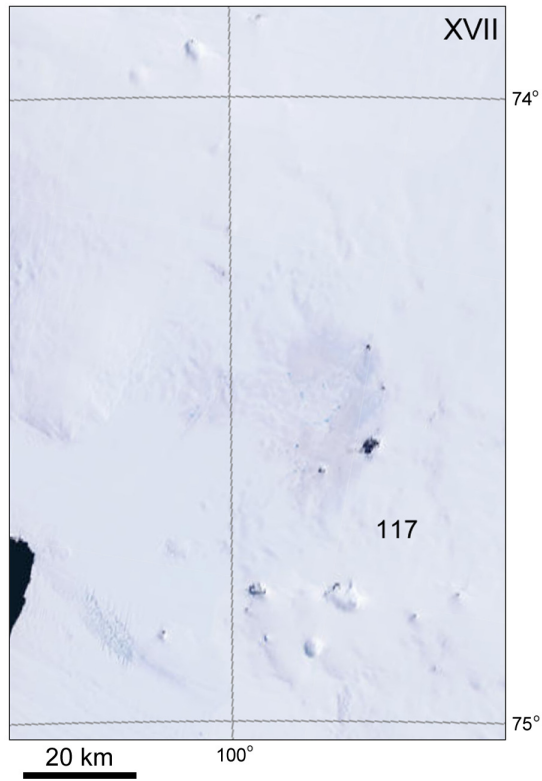

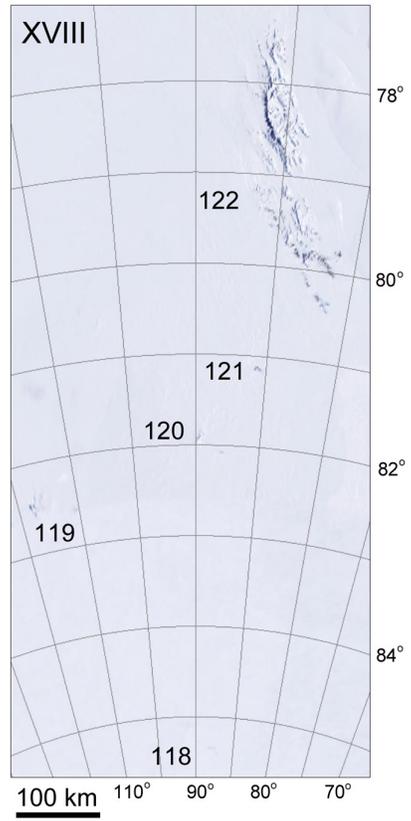

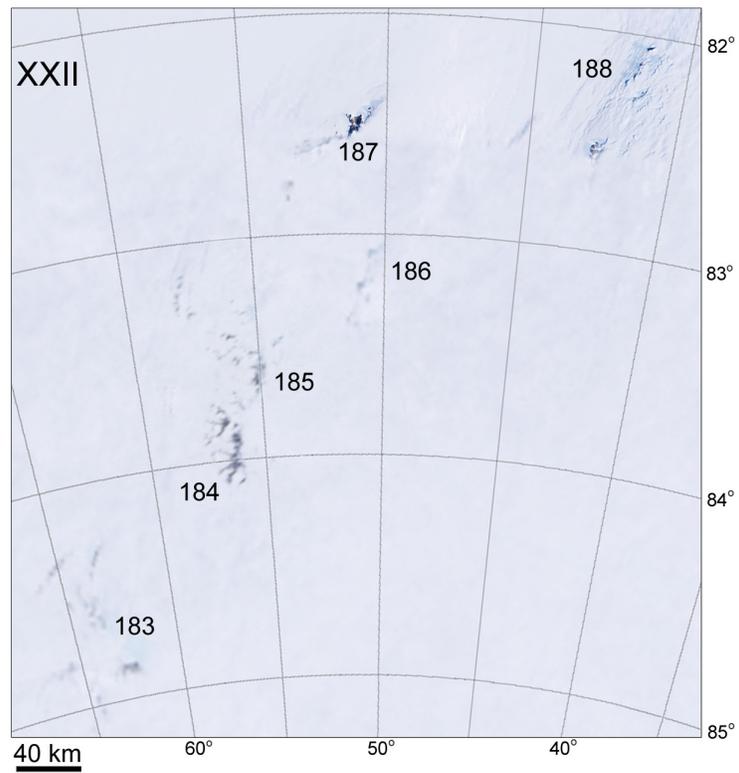

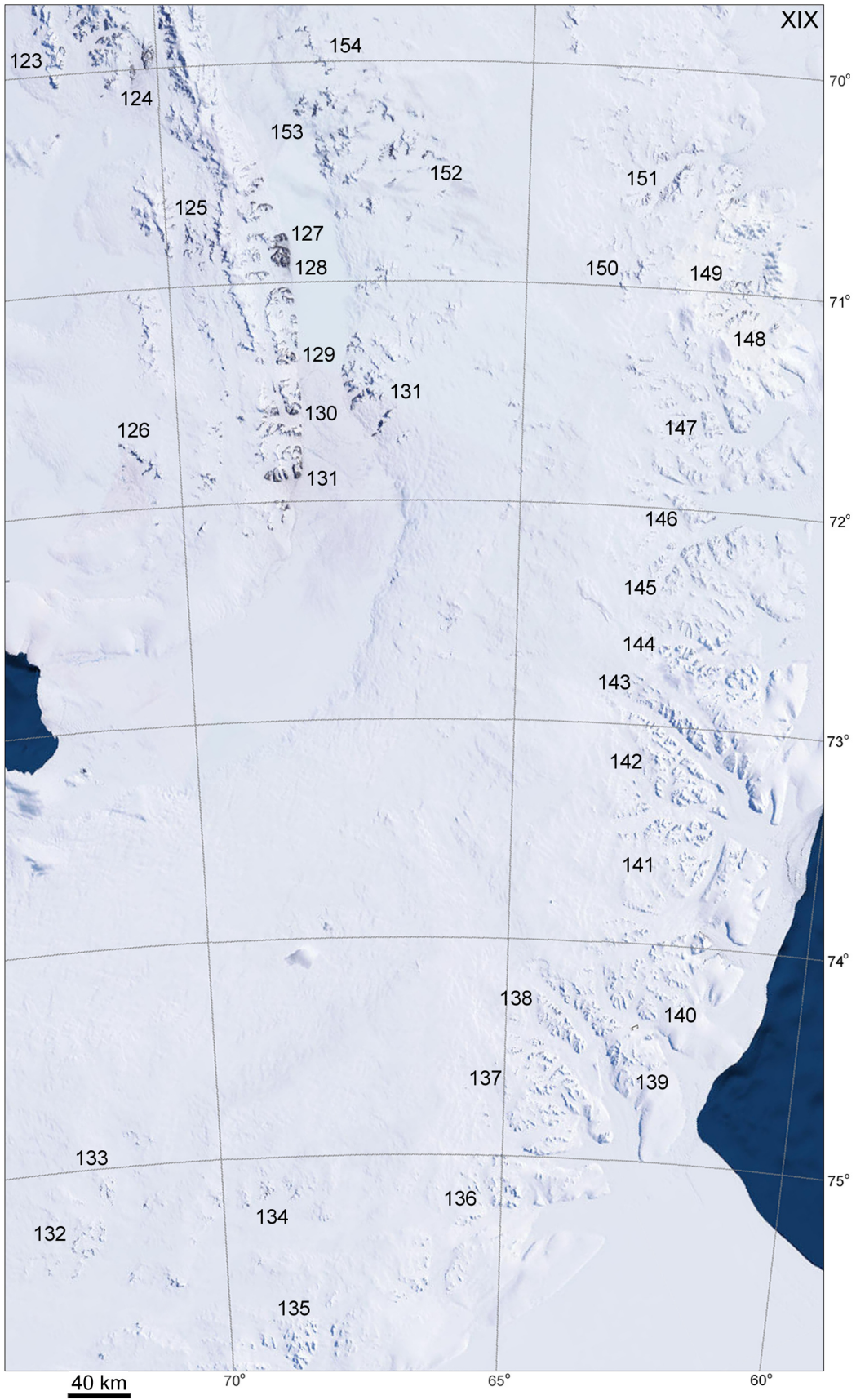

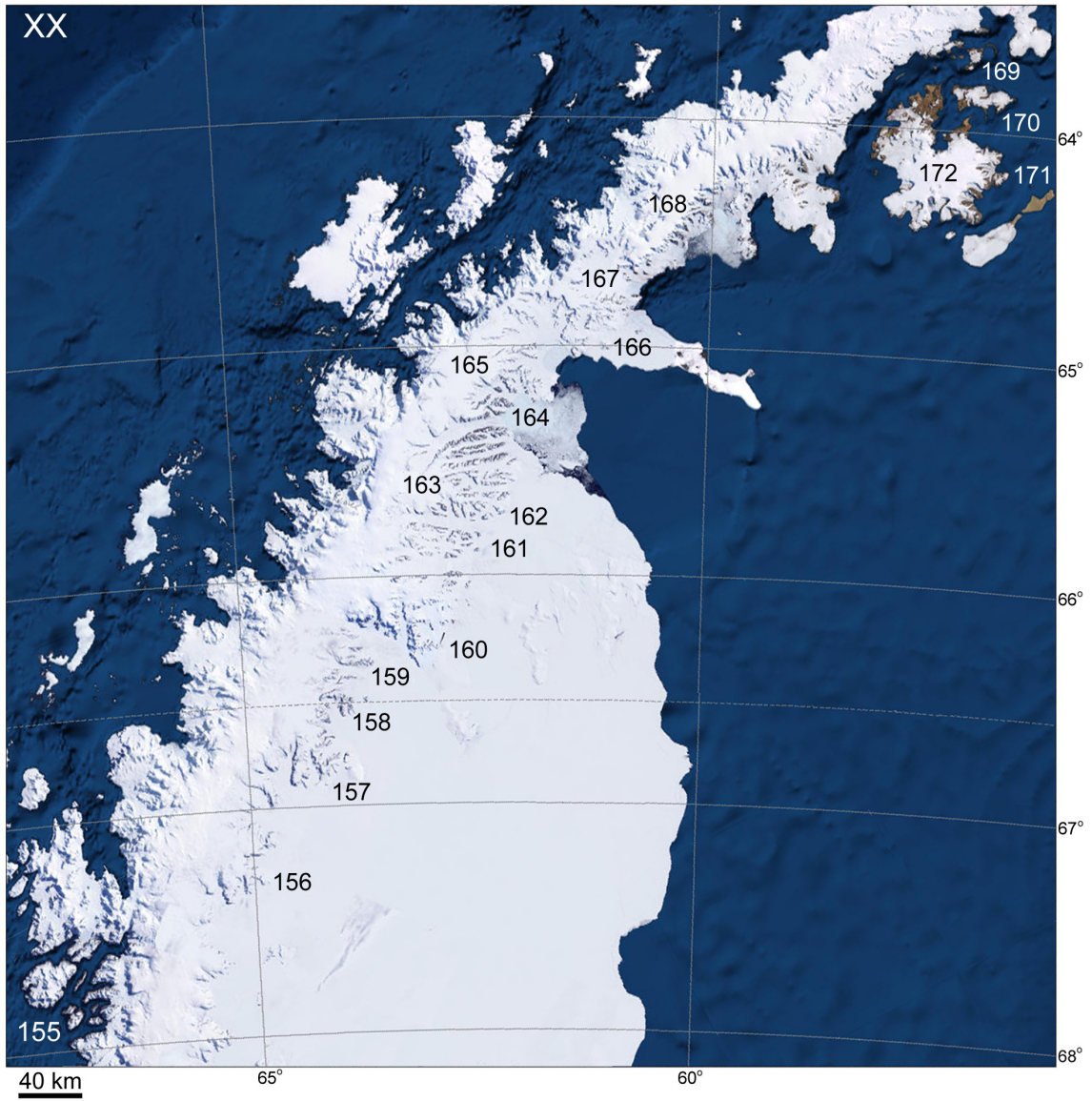

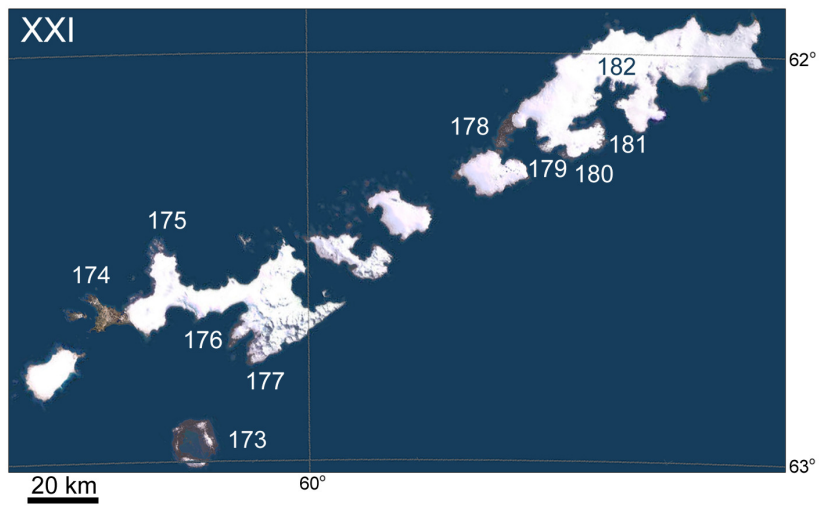

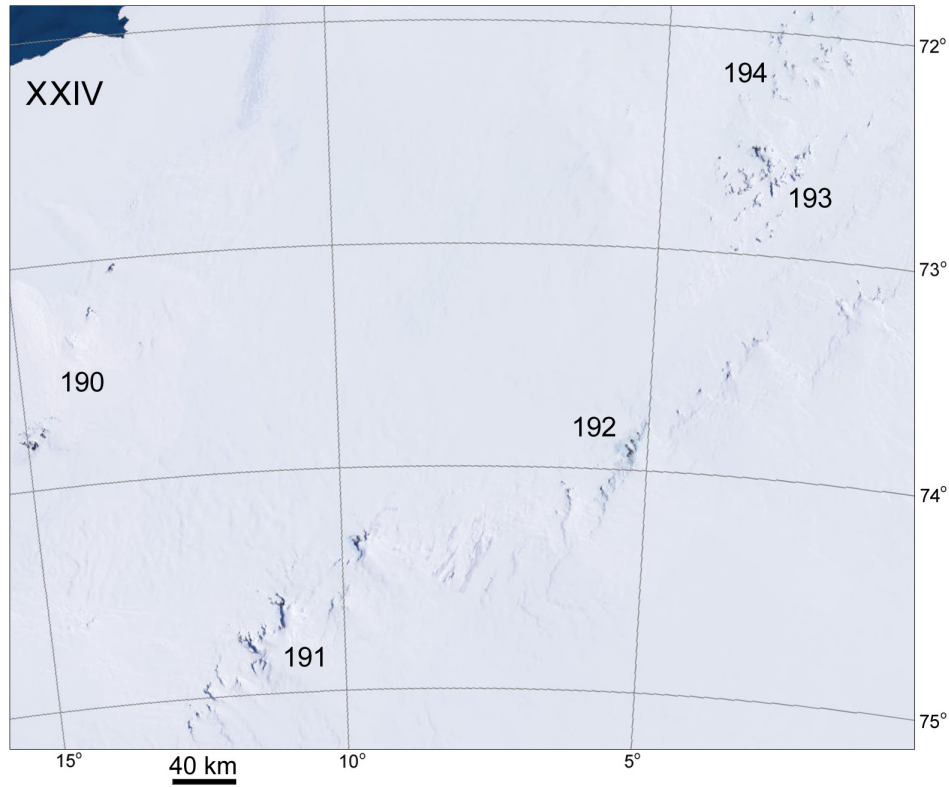

**Fig. 3** Ice-free Antarctic areas. Latin numerals are numbers of satellite imagery maps (see Fig. 2); Arabic numerals are numbers of areas (see Table 1). Satellite imagery mosaics are from the REMA Explorer (PGC 2022–2024); source: Earthstar Geographics

### 2.1.4. Maps, projections, and scales

For each ice-free area, the atlas will contain the following maps and materials:

1.  A hypsometric map.

2.  A series of morphometric maps. The following morphometric attributes will be calculated and mapped: slope gradient, slope aspect, horizontal curvature, vertical curvature, minimal curvature, maximal curvature, catchment area, topographic wetness index, stream power index, and wind exposition index (Table 2).

3.  Textual reference material (explanations for geomorphometric calculations and modeling, physical and geographical interpretations of morphometric maps, etc.).

The following scales and projections will be used in the atlas:

• 1 : 10,000,000 and the polar stereographic projection for one overview hill-shaded map of Antarctic topography.

• 1 : 1,000,000 and Lambert conformal conic projection for one overview hill-shaded map for each chapter of the atlas (that is, for each Antarctic region).

• 1 : 100,000 / 1 : 50,000 / 1 : 25,000 and Universal Transverse Mercator (UTM) projection for a series of morphometric maps for each area depending on the territory size and readability of the derived maps.

The use of three scale levels and projection types will allow the Antarctic topography to be displayed in all its diversity and with varying degrees of detail, from the continental to regional and to local scales.

## 2.2. Data, processing, and mapping
### 2.2.1. Input data

REMA versions 1.1 and 2 with grid spacings of 2, 8, and 10 m (Howat et al. 2022) will be the input data for all geomorphometric calculations, modeling, and atlas mapping. The choice of REMA is due to the fact that currently it is the most complete, detailed, and accurate DEM of Antarctica. REMA was built on the basis of hundreds of thousands of individual local DEMs obtained photogrammetrically using stereo images from the WorldView-1, WorldView-2, and WorldView-3 spacecrafts with a resolution from 0.32 m to 0.5 m. The images were obtained in austral summers of 2009–2021. Each individual DEM was registered vertically using satellite altimetry measurements from the Cryosat-2 and ICESat spacecrafts. As a result, an absolute error of REMA is less than 1 m over most of the coverage area, while its relative errors are in the decimeter range (Howat et al. 2019).

Our previous assessments of the REMA fragments (Florinsky 2023a, 2023b, 2025b, 2025c; Florinsky and Zharnova 2025) showed that accuracy and quality of the REMA coastal portions generally meet the requirements of geomorphometric modeling, namely, low level of high-frequency noise, low number of obvious artifacts, and sufficient surface smoothness not requiring additional smoothing or filtering. However, some local studies have demonstrated that the 2-meter gridded REMA version performs significantly worse in terms of accuracy and artifact-related quality compared to the 8-meter gridded REMA version (Idalino et al. 2021; Florinsky and Zharnova 2025).

As input data for geomorphometric modeling and mapping of all ice-free areas (Table 1), we will use fragments of REMA versions 1.1 and 2 with grid spacings of 2, 8, and 10 m. Each REMA fragment will include one of the ice-free terrains. The choice of specific grid spacing for a particular REMA fragment will depend on the area size, DEM quality (see above), and readability of the derived maps, as in the case of morphometric map scales (Section 2.1). It is important to note that in geomorphometric studies the race for maximum resolution is often unjustified because modeling with less resolution gives similar or even better results (Cavazzi et al. 2013; Chang et al. 2019; Maxwell and Warner 2019; Florinsky 2025a, chap. 21). The spatial resolution used of each ice-free area will be informed for the readers in the final version of the atlas.

As input data for an overview hill-shaded map of Antarctica (1 : 10,000,000 scale) and twenty regional hill-shaded maps (1 : 1,000,000 scale), we will use REMA version 2 with grid spacings of 1000 m (the full model) and 100 m (fragments), respectively.

### 2.2.2. Data pre-processing

Data pre-processing of the REMA fragments will include two main procedures: reprojection and editing.

REMA is presented in the polar stereographic projection; elevations are given relative to the WGS84 ellipsoid (Howat et al. 2019, 2022). For geomorphometric calculations and modeling, 194 REMA fragments will be reprojected into the UTM projection preserving the original grid spacings in interpolation of elevation values. For producing regional hill-shaded maps, 20 REMA fragments will be reprojected into the Lambert conformal conic projection. For all REMA fragments, ellipsoidal elevations will be converted to orthometric ones.

Coastal portions of REMA contain island-like artifacts, which are images of icebergs (Florinsky 2023b; Florinsky and Zharnova 2025). Along the coastline of the South Shetland Islands, REMA contains artifacts like islands, peninsulas, and bars in bays. These are results of incorrect photogrammetric processing of stereo images covering both the land and the sea. To remove the artifacts, we will edit manually the REMA fragments. As reference data, we will use the available topographic maps (Bakaev and Tolstikov 1966; USGS 1959–1973; Korotkevich et al. 2005; AAD 2023), the Antarctic Research Atlas (USGS 1999–2024), and the REMA Explorer (PGC 2022–2024).

### 2.2.3. Calculations and modeling

Digital models of all morphometric variables are calculated from DEMs (Florinsky 2025a, chap. 4).

For each ice-free area (Table 1), we will calculate digital models of a representative set of local, nonlocal, combined, and two-field-specific morphometric variables (or, attributes). A local morphometric variable describes the surface geometry in the vicinity of a given point of the topographic surface. A nonlocal topographic attribute characterizes a relative position of a given point of the topographic surface. A combined morphometric variable can consider both the local geometry of the topographic surface and the relative position of a point of the surface. A two-field-specific morphometric attribute describes relations between the topographic surface located in gravity field and other fields, such as solar irradiation, wind flow, etc. (Florinsky 2017, 2025a, chap. 2).

Digital models of the eleven most scientifically important, fundamental morphometric variables will be derived from the REMA fragments. The list of variables (Table 2) includes six local morphometric attributes: slope gradient ($G$), slope aspect ($A$), horizontal curvature ($k_h$), vertical curvature ($k_v$), minimal curvature ($k_{min}$), and maximal curvature ($k_{max}$); one nonlocal variable—catchment area ($CA$); two combined variables: topographic wetness index ($TWI$) and stream power index ($SPI$); as well as two two-field-specific morphometric attribute—total insolation ($TIns$) and wind exposition index ($WEx$). Formulas and detailed interpretations of these variables can be found elsewhere (Florinsky 2017, 2025a, chap. 2).

The importance of the listed morphometric variables is determined by the following key factors. First, the selected variables describe various features of the topography sufficiently fully to conduct geomorphological studies of various types (Florinsky 2025a, chaps. 1 and 2). Second, the selected morphometric variables most fully describe the dependence of soil/plant properties on topography; that is, these variables are the most effective predictors of soil/plant properties (Hengl and Reuter 2009; Wilson 2018; Florinsky 2025a, chaps. 1 and 9–12). Third, the four curvatures selected are the most informative morphometric variables in geological terms (Florinsky 2025a, chaps. 13–16).

For each ice-free area (Table 1), digital models of all morphometric attributes will be derived from the reprojected and edited REMA fragments. To derive digital models of local morphometric variables (i.e., $G$, $A$, $k_h$, $k_v$, $k_{min}$, and $k_{max}$), we will use the classical finite-difference method by Evans (1980). To compute digital models of $CA$, we will apply the maximum-gradient based multiple flow direction algorithm by Qin et al. (2007) to preprocessed sink-filled DEMs. Since a very wide dynamic range of values characterizes $CA$, its digital models will be logarithmized. To derive digital models of combined morphometric variables (i.e., $TWI$ and $SPI$), we will use previously calculated models of $CA$ and $G$. To calculate digital models of $TIns$ and $WEx$, we will apply two related methods by Böhner (2004). Calculation of the last two morphometric variables requires the choice of some parameters. To derive models of $WEx$, we will select an angular step and a search distance depending on the size of an ice-free area. $TIns$ models will be computed for one summer day (1 January) with a temporal step of 0.5 h.

To produce an overview hill-shaded model of Antarctica and twenty regional hill-shaded models from the 1000-m gridded REMA and 100-m gridded REMA fragments, respectively, we will compute the Lambertian reflectance (Horn 1981) (Table 2). The parameters of this calculation (i.e., solar azimuth and solar elevation angles) will be selected to obtain the most effective hill-shaded maps.

**Table 2 Definitions and interpretations of selected morphometric variables (Shary et al. 2002; Florinsky 2017, 2025a, chap. 2)**

| Variable, notation, and unit | Definition and interpretation |
|---|---|
| *Local morphometric variables* | |
| Slope gradient, $G$ (°) | An angle between the tangential and horizontal planes at a given point of the topographic surface. Relates to the velocity of gravity-driven flows. |
| Slope aspect, $A$ (°) | An angle between the north direction and the horizontal projection of the two-dimensional vector of gradient counted clockwise, from 0° to 360°, at a given point of the topographic surface. Relates to the direction of gravity-driven flows |
| Horizontal curvature, $k_h$ (m$^{-1}$) | A curvature of a normal section tangential to a contour line at a given point of the surface. A measure of flow convergence and divergence. Gravity-driven lateral flows converge where $k_h < 0$, and diverge where $k_h > 0$. $k_h$ mapping reveals crest and valley spurs. |
| Vertical curvature, $k_v$ (m$^{-1}$) | A curvature of a normal section having a common tangent line with a slope line at a given point of the surface. A measure of relative deceleration and acceleration of gravity-driven flows. They are decelerated where $k_v < 0$, and are accelerated where $k_v > 0$. $k_v$ mapping reveals terraces and scarps. |
| Minimal curvature, $k_{min}$ (m$^{-1}$) | A curvature of a principal section with the lowest value of curvature at a given point of the surface. $k_{min} > 0$ corresponds to local convex landforms, while $k_{min} < 0$ relates to elongated concave landforms (e.g., hills and troughs, correspondingly). |
| Maximal curvature, $k_{max}$ (m$^{-1}$) | A curvature of a principal section with the highest value of curvature at a given point of the surface. $k_{max} > 0$ corresponds to elongated convex landforms, while $k_{max} < 0$ relate to local concave landforms (e.g., crests and holes, correspondingly). |
| *Nonlocal morphometric variables* | |
| Catchment area, $CA$ (m$^2$) | An area of a closed figure formed by a contour segment at a given point of the surface and two flow lines coming from upslope to the contour segment ends. A measure of the contributing area. |
| *Combined morphometric variables* | |
| Topographic wetness index, *TWI* | A ratio of catchment area to slope gradient at a given point of the topographic surface. A measure of the extent of flow accumulation. |
| Stream power index, *SPI* | A product of catchment area and slope gradient at a given point of the topographic surface. A measure of potential flow erosion and related landscape processes. |
| *Two-field specific morphometric variables* | |
| Wind exposition index, *WEx* | A measure of an average exposition of slopes to wind flows of all possible directions at a given point of the topographic surface. |
| Total insolation, *Tins* (kWh/m$^2$) | A measure of the topographic surface illumination by solar light flux. Total potential incoming solar radiation, a sum of direct and diffuse insolations. |
| Reflectance, $R$ | A measure of the brightness of an illuminated topographic surface at its given point. |

### 2.2.4. Mapping

Visualization of the overview hill-shaded map of Antarctica and twenty regional hill-shaded maps will be done by the standard procedure of analytical hill shading using the achromatic mode (Jenny 2001).

Almost any REMA fragment includes both an ice-free area and portions of the adjacent continental ice sheet, glaciers, or ice shelves. Therefore, for hypsometric mapping, to ensure adequate perception of the topography, we will use two gradient hypsometric tint scales to depict the relief of each area:

• To display the elevations of the ice-free topography, we will apply the green-yellow part of the standard spectral hypsometric scale of color plasticity (Kovaleva 2014).

• To display the elevations of the glacier topography, we will utilize a modified hypsometric tint scale for polar regions (Patterson and Jenny 2011).

To obtain the final hypsometric map of an ice-free area, two hypsometric tinting will be combined with an achromatic hill-shaded layer derived from the REMA fragments by a standard analytical hill shading (Jenny 2001). The hypsometric maps of ice-free terrains will also show year-round polar stations and seasonal field bases (both operating and closed) as well as airfields, if such objects exist there.

For morphometric mapping of ice-free areas, the following rules will be applied:

• $G$ takes only positive values from 0° to 90°. To map $G$, we will apply a standard gray tint scale (the minimum and maximum $G$ values correspond to white and black, respectively).

• $A$ is circular variable taking values from 0° to 360°. To map $A$, we will apply an eight-color, eight-cardinal direction tint scale.

• $k_h$, $k_v$, $k_{min}$, and $k_{max}$ take both negative and positive values having opposite physical mathematical senses and interpretations (Table 2). To map these variables, we will use a two-color tint scale consisting of two contrasting parts, blue and orange (negative and positive curvature values, respectively). The darkest and lightest shades of blue or orange colors correspond to the absolute maximum and minimum curvature values, respectively.

• $CA$ takes only positive values. To map $CA$, we will apply a standard gray tint scale (the minimum and maximum $CA$ values correspond to white and black, respectively).

• $TWI$ and $SPI$ take only positive values. To map $TWI$ and $SPI$, we will utilize a standard tint scale of spectral colors (the minimum and maximum $TWI$ or $SPI$ values correspond to violet and red, respectively).

• $TIns$ is a nonnegative variable. To map $TIns$, we will apply an orange tint scale: the minimum and maximum $TIns$ values correspond to the darkest and lightest orange shades depicting the least and most illuminated areas, respectively.

• $WEx$ is a positive dimensionless variable, wherein values below and above 1 relate to wind-shadowed and wind-exposed areas, respectively. To map $WEx$, we will use a two-color tint scale consisting of two contrasting parts, orange and violet (values below and above 1, respectively). The darkest shades of orange and violet colors correspond to the minimum and maximum $WEx$ values, respectively, while the lightest shades of the colors correspond to 1.

Unlike the hypsometric map, on the morphometric maps we will not specifically highlight or distinguish the ice-free area from portions of the adjacent ice sheet, glaciers, or ice shelves. Our previous studies have shown that morphometric map patterns for ice-free areas and adjacent ice surfaces differ greatly from each other (Florinsky 2023a, 2023b, 2025b, 2025c; Florinsky and Zharnova 2025). Such a clearly visible contrast is associated, on the one hand, with the smoothness of the glacier surfaces and, on the other hand, with the roughness and dissection of the ice-free surfaces.

There is no lake bathymetry data in REMA, but unlike marine water bodies, lake cells are not marked as "no data". These cells contain interpolated values of lake coastal elevations, that is, artifacts. In this regard, the lakes will be masked on the resulting hypsometric and morphometric maps. As reference data for lakes, we will use the available topographic maps (Bakaev and Tolstikov 1966; USGS 1959–1973; Korotkevich et al. 2005; AAD 2023), the Antarctic Research Atlas (USGS 1999–2024), and the REMA Explorer (PGC 2022–2024).

Figures 4–6 show examples of morphometric maps for four ice-free areas, such as the Thala Hills (Molodezhny and Vecherny Oases), Schirmacher Hills, and Fildes Peninsula. Complete series of morphometric map for the ten periglacial areas—the Larsemann Hills,

Thala Hills (Molodezhny and Vecherny Oases), Schirmacher Hills, Fildes Peninsula, Bunger Hills, Cape Burks, Haswell Island, Leningradsky Nunatak, and Gaussberg Volcano—can be found elsewhere (Florinsky 2023a, 2023b, 2025b, 2025c; Florinsky and Zharnova 2025).

### 2.2.5. Software

For reprojecting and editing the REMA fragments as well as geomorphometric calculations, modeling, and mapping, we will use the SAGA (Conrad et al. 2015) and ArcGIS Pro (ESRI 2015–2024) software, which have already proven themselves well for these purposes (Florinsky 2023a, 2023b, 2025b, 2025c; Florinsky and Zharnova 2025).

### 2.2.6. Atlas publication

We plan to publish the atlas as a 400-pages volume of the folio size. In addition, all derived digital morphometric models will be stored online as GEOTIFF files. They will be available for free downloading. This will give Antarctic researchers an opportunity to use the obtained geomorphometric information in their studies.

## 3. Discussion

### 3.1. Specificity of ice-free Antarctic areas

The idea of creating geomorphometric atlases has previously been expressed several times (Guth 2007; Csillik and Drăguţ 2018). The first successful attempt was the geomorphometric atlas of Romania implemented in the form of a geoportal (Ioniţă et al. 2024). On the one hand, the scientific community perceives the idea of regional geomorphometric atlases positively. On the other hand, this field of atlas mapping has not yet been sufficiently developed. Therefore, in the process of creating the atlas, some methodological tasks and questions may arise.

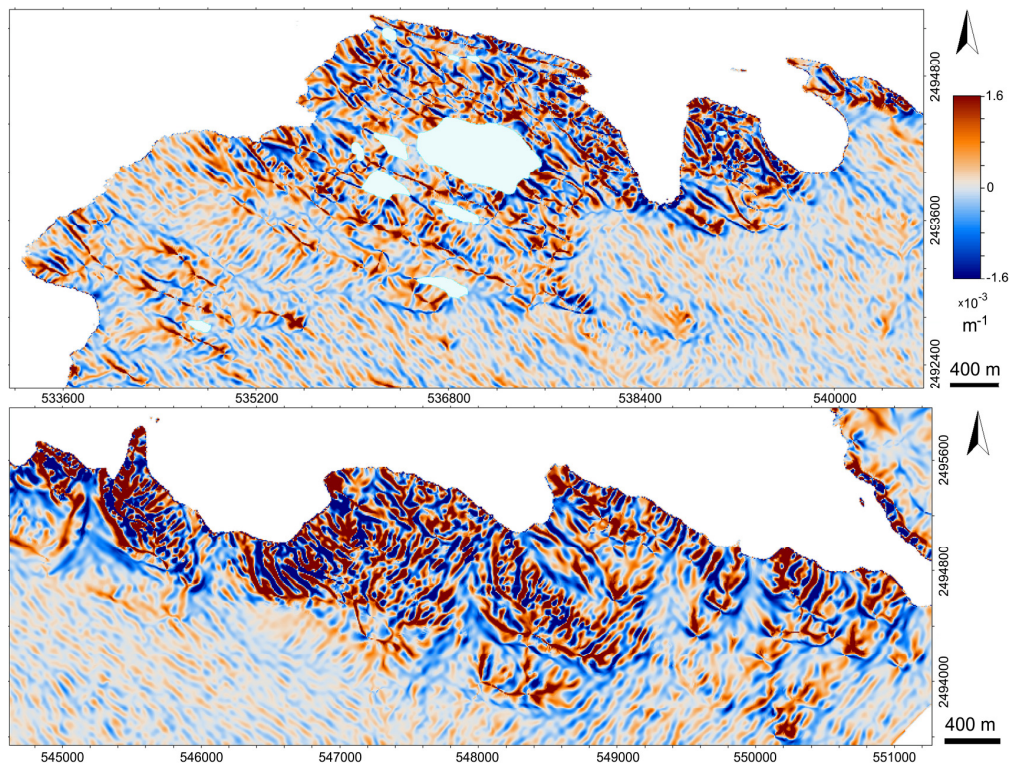

**Fig. 4** Thala Hills, Enderby Land, East Antarctica; Molodezhny and Vecherny Oases (upper and lower, respectively): horizontal curvature. Light blue areas are lakes (Florinsky 2023a)

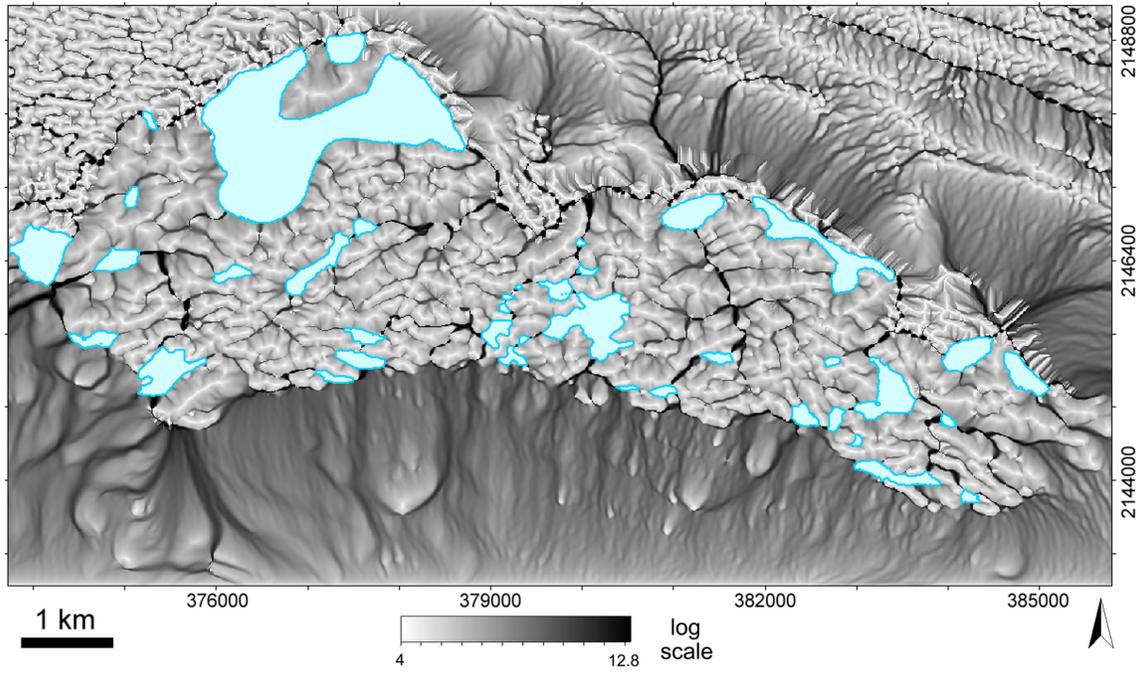

**Fig. 5** Eastern part of the Schirmacher Hills, Queen Maud Land, East Antarctica: catchment area. Light blue areas are lakes (Florinsky 2023a)

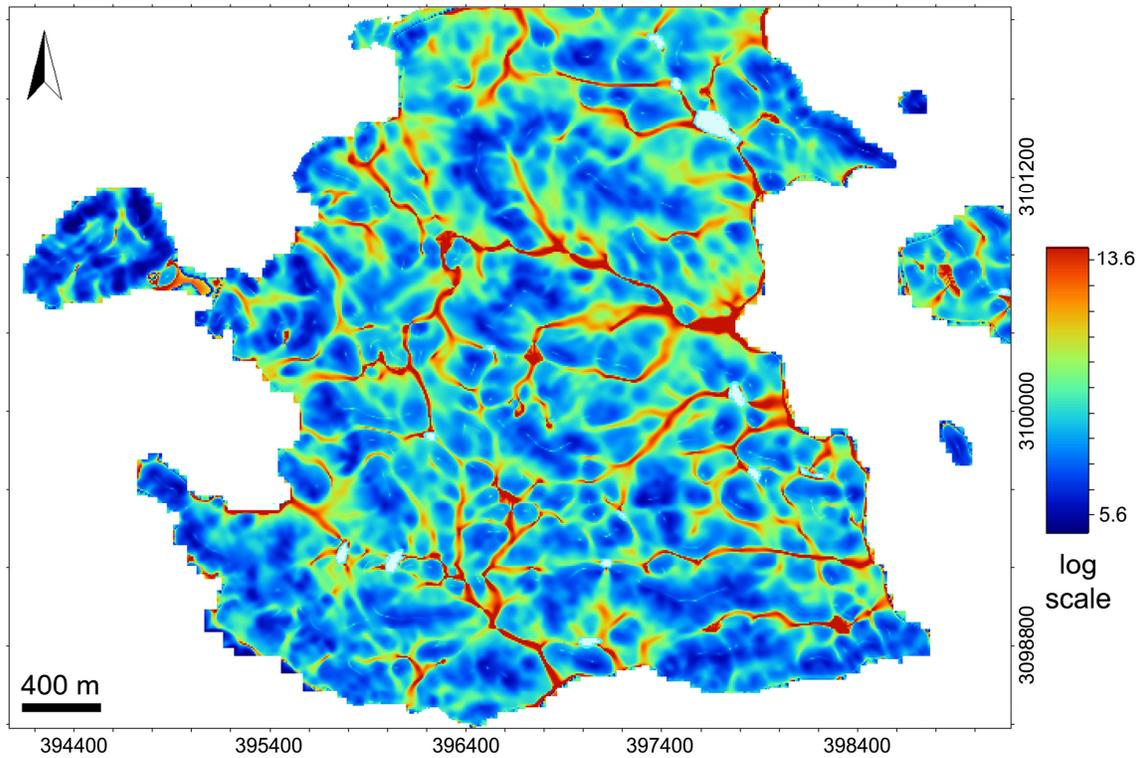

**Fig. 6** Southern part of the Fildes Peninsula, King George Island, South Shetland Islands: topographic wetness index. Light blue areas are lakes (Florinsky 2023a)

Moreover, three specific features of the periglacial Antarctic areas pose particular challenges in developing the atlas:

1. A wide range of sizes of such terrains. This feature is clearly visible in a comparative analysis of the size of ice-free areas within the same satellite imagery map. For example, on the satellite imagery map I showing Queen Maud Land (Fig. 3), one can see that the Wohlthat Mountains (# 9) and the Schirmacher Hills (# 14) have sizes of about 110 km × 20 km and 20 km × 3 km, respectively. As a result, within the frames of a single-volume atlas with a relatively limited size, even neighboring objects will be represented on maps of essentially different scales (e.g., 1 : 100,000 vs. 1 : 25,000 for the Wohlthat Mountains and the Schirmacher Hills, respectively).

2. Uneven spatial distribution of ice-free areas. For example, let us compare the satellite imagery maps XIII and VIII (Fig. 3). The map XIII (Victoria Land) shows the McMurdo Dry Valleys (# 82) stretching from north to south for about 170 km. The map VIII (William II Land – Queen Mary Land) shows two objects—Mount Gaussberg (# 58) and the Haswell Islands (# 59)—the distance between which is also about 170 km. In Adélie Land, there is no ice-free area large enough to be displayed in the atlas. Thus, representation of different Antarctic regions in the atlas will vary greatly.

3. Uncertainty of the boundaries of ice-free areas. The issue of determining the boundaries of Antarctic oases has earlier been raised in the context of their landscape mapping (Alexandrov 1985). Indeed, is it permissible to draw the boundary of an ice-free area formally along its perimeter if these boundaries include small snowfields or passive ice domes located inside the oasis? How to draw the boundaries of a partially ice-free mountain area if the ridges are free of ice, but the valleys between them are not? Should nunataks located nearby be included in the oasis, or should they be considered as separate ice-free objects? This challenge can be exemplified by the case of the Molodezhny and Vecherny Oases (Fig. 3, satellite imagery map III; ## 27 and 28, respectively). Although there is approximately 20 km of ice between these two areas, some researchers consider them as a single oasis, the Thala Hills. In general, the problem of territory boundaries will be solved for each area individually when creating its morphometric maps; apparently, there is no a universal and unbiased solution to this problem.

### 3.2. Key properties of the atlas

Any atlas should possess integrity, which is realized due to two mandatory key properties of an atlas: the completeness and internal unity (Salichtchev 1990, pp. 186–187).

The completeness of the geomorphometric atlas will be ensured by two factors:

1. The atlas will present the vast majority of ice-free areas of Antarctica.

2. For these terrains, the atlas will include maps of all key, scientifically important, and clearly interpretable morphometric attributes.

The internal unity of the geomorphometric atlas will be ensured by three factors:

1. A series of morphometric maps of each area will be created at the same scale, which will ensure ease of comparison of these maps.

2. A series of maps of one representative set of fundamental attributes from the theory of the topographic surface and the concept of general geomorphometry will be calculated for all areas. This will facilitate a comparative analysis of the terrains located in different regions with various geological and geomorphic conditions.

3. Each morphometric map quantitatively describes a certain property of the topographic surface and has a unique physical mathematical and physical geographical interpretation. At the same time, morphometric maps of one series complement each other.

The principal feature and novelty of the atlas will be the system view on maps of key fundamental morphometric attributes associated with the theory of the topographic surface and the concept of general geomorphometry (Section 2.1.1).

### 3.3. Possible applications of the atlas

The geomorphometric information to be presented in the atlas is important not only because it is a new quantitative knowledge about the ice-free Antarctic topography. It can also be used in soil, ecological, geological, and other types of research. Let us consider some possible options for the further use of morphometric data from the atlas.

It is known that topography controls the thermal, wind, and hydrological regimes of slopes, influencing therefore the distribution and properties of soils and vegetation (Huggett and Cheesman 2002; Florinsky 2025a).

The thermal regime of slopes depends in part on the incidence of solar rays to the land surface, so it depends on both $G$ and $A$. Insolation $TIns$ directly describes this incidence and therefore better considers the thermal regime (Böhner and Antonić 2009). Information on the differentiation of slopes by insolation level is critical to predict the spatial distribution of primitive soils and lower plants in the ice-free areas. To refine such a prediction, one can use $TWI$ digital models describing topographic prerequisites of water migration and accumulation.

A greater refinement of such a prediction is possible by including $WEx$ data in the analysis. $WEx$ digital models can be utilized to identify areas affected by and protected from the wind impact (Böhner and Antonić 2009; Florinsky 2025a, chap. 2). In periglacial Antarctic landscapes, one of the main meteorological factors determining the microclimate is katabatic wind. Thus, $WEx$ digital models may be of great importance for modeling the distribution of primitive soil and lower plants, predicting the differentiation of snow accumulation in ice-free areas, and determining the optimal location of buildings and infrastructure at polar stations.

$k_h$ maps display the distribution of convergence and divergence areas of surface flows. Geomorphologically, these are spurs of valleys and crests, respectively. The combination of convergence and divergence areas creates an image of the flow structures (Florinsky 2025a, chap. 2). $k_v$ maps show the distribution of relative deceleration and acceleration areas of flows. Geomorphologically, these maps represent cliffs, scarps, terrace edges, and other similar landforms or their elements with sharp bends in the slope profile (Florinsky 2025a, chap. 2). In this regard, $k_h$ and $k_v$ digital models may be useful in geomorphological and hydrological studies of the ice-free areas.

Combination of $k_h$ and $k_v$ digital models allows revealing relative accumulation zones of surface flows (Florinsky 2025a, chap. 2). These zones, marked by both $k_h < 0$ and $k_v < 0$, coincide with the fault intersection sites and are characterized by increased rock fragmentation and permeability. Within these zones, one can observe an interaction and exchange between two types of substance flows: (a) lateral, gravity-driven substance flows moved along the land surface and in the near-surface layer, such as water, dissolved and suspended substances, and (b) vertical, upward substance flows, such as fluids, groundwater of different mineralization and temperature (Florinsky 2025a, chap. 15). Maps of accumulation zones may be useful for geochemical studies in the ice-free areas. An example of such a map for the Larsemann Hills can be found elsewhere (Florinsky 2023b).

$k_{max}$ and $k_{min}$ maps are informative in terms of structural geology because they display elongated linear landforms (Florinsky 2017, 2025a). In Antarctica, such lineaments can be interpreted as a reflection of the local fault and fracture network, which topographic manifestation has been amplified by erosional, exaration, and nival processes (Florinsky 2023b; Florinsky and Zharnova 2025). Thus, researchers will be able to use $k_{max}$ and $k_{min}$ data for compiling lineament maps and comparing them with other geological data. An example of such a lineament map for the Larsemann Hills can be found elsewhere (Florinsky 2023b).

$CA$ maps can be used to identify the fine flow structure of drainage basins and then to

incorporate this information into geochemical and hydrological analysis. *TWI* digital models can be used for spatial prediction of the ground moisture content in the ice-free areas as well as for forecasting the spatial distribution of snow puddles on adjacent glaciers in summer. *SPI* data can be useful for spatial prediction of slope erosion in the ice-free areas as well as erosion of snow cover and ice by meltwater flows on adjacent glaciers in summer.

## 4. Conclusions

We rationalized the need to develop a physical geographical, scientific reference geomorphometric atlas of ice-free Antarctic areas. We presented the concept, structure, and content of the atlas including the list of areas to be modeled and mapped. Data and methods of geomorphometric calculation, modeling, and mapping are described in detail.

The atlas will possess integrity due to its completeness and internal unity. However, its principal feature and novelty will be the system view on maps of key fundamental morphometric attributes associated with the theory of the topographic surface and the concept of general geomorphometry. Each morphometric map quantitatively describes a certain property of the topography and has a unique physical mathematical and physical geographical interpretation. Maps of various morphometric characteristics complement each other.

The atlas will concentrate multi-scale, multi-aspect quantitative information on the ice-free Antarctic topography, will present it in a systematized, organized, and easy-to-read form as well as will provide scientific and information support for fundamental and applied research in Antarctica.

The project is expected to be completed by 2030.